\documentclass[11pt]{article}
\usepackage{amsmath}
\usepackage{amssymb}

\makeatletter
\usepackage{latexsym}
\usepackage{amsfonts}

\@addtoreset{equation}{section}

\newcommand{\eref}[1]{(\ref{#1})}

\def\be{\begin{equation}}
\def\ee{\end{equation}}
\def\ba{\begin{eqnarray}}
\def\ea{\end{eqnarray}}
\def\bet{\begin{tabular}}
\def\eet{\end{tabular}}
\def\pa{\partial}

\def\nn{\nonumber}
\def\ve{\varepsilon}

\def\a{\alpha}

\def\dl{\delta}
\def\la{\lambda}

\def\m{\mu}
\def\n{\nu}
\def\ra{\rightarrow}
\def\vp{\varphi}

\long\def\symbolfootnote[#1]#2{\begingroup%
\def\thefootnote{\fnsymbol{footnote}}\footnote[#1]{#2}\endgroup}

\makeatother

\begin{document}
\begin{titlepage}\global\long\def\LimV{\underset{V\rightarrow1}{\mathrm{Lim}}\;}
\global\long\def\LimE{\underset{\varepsilon\rightarrow0}{\mathrm{Lim}}\;}

\begin{flushright}
Preprint DFPD/2013/TH22\\
 January 2014\\

\par\end{flushright}

\vspace{0truecm}

\begin{center}
\textbf{\Large Electromagnetic fields and potentials generated by
massless charged particles \vskip0.3truecm}{\Large{} }
\par\end{center}{\Large \par}

\begin{center}
\vspace{0.2cm}

\par\end{center}

\begin{center}
Francesco Azzurli$^{1}$\symbolfootnote[1]{francesco.azzurli@gmail.com}
and Kurt Lechner$^{2}$\symbolfootnote[2]{kurt.lechner@pd.infn.it}
\par\end{center}

\begin{center}
\vspace{1.2cm}
 $^{1}$\textit{Scuola Galileiana di Studi Superiori,  Universit\`a
degli Studi di Padova, Italy}
\par\end{center}

\begin{center}
%\vspace{0.5cm}
 $^{2}$\textit{Dipartimento di Fisica e Astronomia, Universit\`a degli
Studi di Padova, Italy}
\par\end{center}

\begin{center}
\textit{and}
\par\end{center}

\begin{center}
\textit{
%\smallskip{}
 INFN, Sezione di Padova,}
\par\end{center}

\begin{center}
\textit{Via F. Marzolo, 8, 35131 Padova, Italy}
\par\end{center}

%\begin{center}
\vspace{0.3cm}
%\par\end{center}

\begin{abstract}
\vskip0.2truecm

We provide for the first time the exact solution of Maxwell's equations for a
massless charged particle moving on a generic trajectory at the speed of light.
In particular we furnish explicit expressions for the vector potential and the electromagnetic field, which were both previously unknown,
finding that they entail different physical features for bounded and unbounded trajectories. With respect to the standard Li\'enard-Wiechert field the electromagnetic field acquires singular $\delta$-like contributions whose support and
dimensionality depend crucially on whether the motion is a)
linear, b) accelerated unbounded,  c) accelerated bounded.
In the first two cases the particle generates a planar shock-wave-like
electromagnetic field traveling along a straight line. In the second and third
cases the field acquires, in addition, a $\delta$-like
contribution supported on a physical singularity-string attached to the particle.
For generic accelerated motions a genuine radiation field is also present,
represented by a regular principal-part type distribution diverging
on the same singularity-string.

\vspace{0.1cm}

\end{abstract}
\end{titlepage}

\newpage

\baselineskip 5mm

\section{Introduction}

Massless particles are rare in nature, the only -- observable and observed --
free massless particle being the photon. In particular \textit{charged}
massless particles do not seem to exist at all, although apparently
there is no fundamental theoretical principle that prohibits their
existence. Actually in general the presence of a massless particle
in classical as well as quantum theories gives rise to \textit{infrared}
long range -- or equivalently low energy -- singularities, that may
undermine the consistency properties of the theory itself. Nevertheless
in quantum field theory, according to the Kinoshita-Lee-Nauenberg
theorem \cite{K,LN}, collinear infrared divergences caused by massless
charged particles, that plague \textit{a priori} the transition amplitudes,
cancel from cross sections if an appropriate sum over degenerate external
initial and final states is performed. There are, however, still open
questions regarding the complete cancellation of these divergences related
to the initial states \cite{LM}, and it is still unclear whether or
not the traditionally considered set of states furnishes a complete
set of physical observables \cite{W}.

In absence of a clear-cut answer regarding the possible existence
of massless charged particles in the framework of quantum field theory,
in this paper we examine the question of the theoretical consistency
of such particles from the point of view of classical Electrodynamics,
addressing the problem of exact solutions of Maxwell's equations.
Obviously the consistency of a classical theory does in general not
imply the consistency of the related quantum theory, and vice-versa,
but a profound analysis of the former may entail a better understanding
of the latter.

Solving Maxwell's equations $\eqref{MaxEq}$ for a charged \textit{point-like}
particle,  in which case the current \eref{LLCurrent} involves a $\delta$-function,
amounts to solve those equations in the \textit{distributional sense} --
the only framework where they make sense. For massive particles,
or equivalently for time-like trajectories, the solution of this problem
has been provided long ago by A.M. Li\'enard and E.J. Wiechert, while
for massless particles, or equivalently for light-like trajectories,
due to the aforementioned singularities it is almost impossible to
solve Maxwell's equations relying on conventional techniques, as the
\textit{Green function method}, see Section \ref{GreenFail}. The Green function method can fail in two respects: {\it i)} because the convolution between two distributions is not defined; {\it ii)} because the formal solution -- represented by the convolution -- does not satisfy the equation one wants to solve, although the Green's function satisfies the proper equation.

In absence of standard tools for solving partial differential equations
we will rely on two different limiting procedures in the space of distributions.
A massless charged particle travels at the speed of light and,
as long as its energy is not directly involved in the physical process
one considers, it appears natural to regard it as the limiting case
of a particle traveling along a time-like \textit{regularized} trajectory
at a speed $V<1$. With this respect a speed $V<1$ in some
sense plays a role similar to the mass $m>0$ used frequently in quantum
field theory to regularize infrared divergences. For a time-like trajectory
the potentials and fields are the ones derived by Li\'enard and Wiechert
and accordingly our first limiting procedure consists in deriving
the potentials and fields we search for, through appropriate distributional
limits of the formers as $V\ra 1$. This procedure entails three crucial ingredients:
a) the proof that the envisaged limits exist and that the corresponding potentials and fields represent, hence, well-defined distributions; b) the proof that the so derived fields satisfy Maxwell's equations; c) the explicit evaluation of the fields.
As we will see, the advantage of this limiting procedure
is that it is \textit{universally} applicable, while its main -- but
merely technical -- drawback is that at intermediate stages it breaks
\textit{manifest} Lorentz-invariance, although it leads to manifest
Lorentz-invariant results.

Our second distributional limiting procedure relies on a {\it manifestly}
Lorentz-invariant regularization of the massless Green function, see
$\eqref{Geps}$, and involves the same three steps mentioned
above. The main advantage of this procedure is obviously its manifest
Lorentz-invariance at all stages, while its principal drawback is
its reduced applicability, in that it works only for \textit{bounded}
trajectories. A part from this the two procedures entail different
technical advantages but, most importantly, when both can be applied,
for uniqueness reasons they lead to the same results.

Due to their conceptual and practical relevance in Electrodynamics,
especially for what concerns the solution of the Bianchi identity
$\partial_{[\alpha}F_{\beta\gamma]}=0$ and the reduction of Maxwell's
equations to the simplified equation $\square A^{\mu}=J^{\mu}$, throughout
the paper we will devote particular attention to the construction
of \textit{potentials}, a delicate issue since -- due to gauge invariance
-- contrary to the fields they can, and will, entail {\it unphysical} singularities. Indeed -- contrary to what is often stated in the literature --
even the limit for $V\ra 1$ of the simple
Li\'enard-Wiechert potential $\eqref{AURV}$ of  a {\it uniform linear} motion with speed $V<1$,  is not a distribution and does not satisfy Maxwell's equations, although
the corresponding Li\'enard-Wiechert \textit{field} admits a well-defined distributional
limit satisfying them (see Section \ref{limAUR}). Nonetheless the construction
of well-defined potentials, although subtle, is an extremely fruitful
task, since it simplifies extremely the construction of well-defined
fields, reducing it essentially to the evaluation of distributional
derivatives.

The results of this paper can be summarized as follows. We prove that
Maxwell's equations for light-like trajectories admit unique causal
solutions in the sense of distributions and we determine
the resulting electromagnetic fields analytically. These fields entail qualitatively
different, in some sense unexpected, expressions according to the qualitative form of the trajectory, involving in any case Dirac $\delta$-like contributions
supported on curves or surfaces. For a \textit{linear} motion (the unique exactly solved case available in the literature \cite{RR,AE}, see also \cite{JKO})
the field is a $\delta$-like planar shock-wave traveling at the speed of
light together with the particle. For accelerated motions the field
becomes singular along a kind of {\it observable} Dirac-like string $\gamma$,
with one endpoint attached to the particle. For accelerated {\it bounded}
motions the second endpoint of $\gamma$ stays at infinity and the
field is made out of two terms, each carrying exactly \textit{one
half} of the electric flux: the first term is a principal-part type ``regular''
distribution diverging on $\gamma$, representing the radiation field,
and the second term is a $\delta$-distribution supported on $\gamma$.
For accelerated {\it unbounded} motions, that we choose to tend
for $t\rightarrow-\infty$ asymptotically to a straight line ${\cal L}$,
the second endpoint of $\gamma$ ends on ${\cal L}$ and -- apart
from the two terms appearing for bounded motions -- the field acquires
an additional term represented by a $\delta$-like shock-wave traveling
along ${\cal L}$ at the speed of light, as if it were due to a
{\it virtual} massless particle in linear motion along ${\cal L}$. This last feature may appear
rather unexpected and we will return to its physical interpretation
in the concluding Section \ref{conclusions}. As anticipated above,
for each kind of trajectory we derive distribution valued well-defined
four-potentials, from which we recover the fields.

The paper is structured as follows. In Section \ref{urm} we illustrate
the failure of the Green function method for massless charges and
apply our first limiting procedure to a {\it linear} motion. It turns out
that even in this simple case the Li\'enard-Wiechert potential admits a limit for a light-like trajectory only after a suitable gauge transformation. In Section \ref{bm} we derive potentials and fields for bounded motions, relying on the covariant limiting procedure. In Section \ref{LWreg} -- that creates a bridge between bounded and unbounded trajectories -- we introduce the alternative Li\'enard-Wiechert limiting procedure and illustrate its drawbacks and virtues again in the case of bounded trajectories. This section contains also the derivation of the distributional limit for $V\ra 1$ of a certain current $K^\m$ -- the source of the Li\'enard-Wiechert radiation field -- that will play a crucial role in
the subsequent Section \ref{ubm}. In this final section we consider
unbounded accelerated trajectories and rely now, for the reasons explained
above, on the Li\'enard-Wiechert limiting procedure to construct potentials.
The analysis of this case is more complex, first because the regularized potential, as for a linear motion, admits a limit only after a suitable gauge transformation, and second, because of the appearance of an unexpected shock-wave along the asymptotic straight line ${\cal L}$ mentioned
above. The concluding Section \ref{conclusions} is devoted to the
physical interpretation of our results and to outlooks. More
involved proofs and derivations are relegated to appendices, Section \ref{app}.

The present paper furnishes in particular the proofs not given in
\cite{AL}.

\section{Li\'enard-Wiechert fields and linear light-like trajectories} \label{urm}

Before facing the solution of Maxwell's equations for light-like trajectories we recall some basic facts about the standard Li\'enard-Wiechert fields and potentials of time-like trajectories, that will play a crucial role in the derivation of the electromagnetic field for light-like trajectories as well.

\subsection{Li\'enard-Wiechert fields and potentials}

We consider a generic {\it time-like} world-line $Y^{\mu}(\lambda)$
and denote its four-velocity and four-acceleration respectively by $U^{\mu}(\lambda)= dY^{\mu}/d\lambda$ and $W^{\mu}(\lambda)= dU^{\mu}/d\lambda$. We denote the three-dimensional  velocity with $\vec V=\vec U/U^0= d\vec Y/dt$ and we will also use the notation ${\mathcal V}^\m=dY^\m/dt=  (1,\vec V)$.
Maxwell's equations for a point particle with charge $e$ read then
\begin{equation}
\partial_{\mu}F^{\mu\nu}= e\!\int\!\delta^{4}(x-Y(\lambda))\,dY^\m\equiv J^{\mu},\quad \quad\partial_{[\m}F_{\n\rho]}=0.\label{MaxEqV}
\end{equation}
We have chosen to define the four-velocity $dY^\m/d\la$ as the derivative of $Y^\m$ with respect to an arbitrary parameter $\la$, since for light-like trajectories, to be considered below, the proper time parameter $s$ is not defined. Due to the appearance of the $\dl$-function in the current $J^\m$,  the solution of Maxwell's equations can be faced consistently only in the {\it space of distributions}. In what follows we will employ the space of {\it tempered} distributions ${\cal S}'({\mathbb R}^4)\equiv \mathcal{S}'$ (see Appendix I for some details).

Solving the Bianchi identity in \eref{MaxEqV} through $F^{\m\n}=\pa^\m A^\n-\pa^\n A^\m$, and resorting to the Lorenz gauge $\partial_{\mu}A^{\mu}=0$, the system $\eqref{MaxEqV}$ is equivalent to
\begin{equation}
\square{A}^{\mu}={J}^{\mu}.\label{boxAUR}
\end{equation}
Relying on the {\it Green function method} this equation entails the standard (retarded)  causal  solution\footnote{Even for time-like trajectories {\it a priori} the convolution in \eref{greenEq} -- between two distributions -- is not well-defined.}
\begin{equation}
{A}^{\mu}=G*{J}^{\mu},\quad\quad G(x)=\frac{1}{2\pi}\,H(x^{0})\,\delta(x^{2}),\quad\quad\square G(x)=\delta^{4}(x),\label{greenEq}
\end{equation}
where $H$ denotes the Heaviside function.
Proceeding formally one gets the Li\'enard-Wiechert potential ($(UL)\equiv U^\m L_\m$ {\it etc}.)
\begin{alignat}{1}
{A^{\mu}}(x)=G*{J}^{\mu} & =\frac{e}{2\pi}\int \!H(x^{0}-\xi^{0})\,\delta\!\left((x-\xi)^{2}\right)\delta^{4}(\xi-Y(\lambda))U^{\mu}(\la) \, d^{4}\xi\, d\lambda \label{convAUR}\\
 & =\frac{e}{2\pi}\int\! H(x^{0} -Y^0(\la))\,\delta\!\left((x- Y(\lambda))^{2}\right) \! U^{\mu}(\la) \, d\lambda\label{AV0}\\
  &=\frac{e}{4\pi} \left.\frac{U^{\mu}}{(UL)}\right|_{\lambda=\lambda(x)},\label{AV}
 \end{alignat}
where $\lambda(x)$ is the retarded parameter determined uniquely by the relations
\begin{equation}
\left(x-Y(\lambda)\right)^{2}=0,\quad\quad x^0\ge Y^{0}(\lambda),\label{lret}
\end{equation}
and
\begin{equation}
L^{\mu}\equiv x^{\mu}-Y^{\mu}(\lambda(x)),\quad\quad L^{2}=0.\label{LV}
\end{equation}
In the following it is understood that the kinematic variables
$Y^{\mu}$, $U^{\mu}$ and $W^{\mu}$ are evaluated at $\lambda(x)$.

It is easy to check that for time-like world-lines the potential \eref{AV} constitutes a distribution. To this order  we apply it to complex test function $\vp(x)$ belonging to the Schwarz space $ {\cal S}({\mathbb R}^4)\equiv {\cal S}$ (see Appendix I). From the equivalent form \eref{AV0}, integrating over $x^0$ and parameterizing the world-line with time, $Y^0(\la)= \la$, we obtain
\begin{alignat}{1}
{A}^{\mu}(\varphi)&=\frac{e}{2\pi}\!\int\! H(x^{0}-\la) \, \delta\!\left((x-Y(\la))^{2}\right)U^\m(\lambda)\,\varphi(x)\,d^4x\,d\la\label{ABconv}\\
&=\frac{e}{4\pi}\int
\frac{{\mathcal V}^{\mu}(\la)}{\big|\vec{x}-\vec{Y}(\la)\big|}\,\varphi\big(\la+
|\vec{x}-\vec{Y}(\la)|,\vec{x}\,\big)\,d^3x\,d\la.
\end{alignat}
Through the shift $\vec{x}\rightarrow\vec{x}+\vec{Y}(\la)$ this integral
becomes eventually
\begin{equation}
{A^{\mu}}(\varphi)=\frac{e}{4\pi}\!\int\frac{{\mathcal V}^{\mu}(t)}{r}\,
\varphi\big(t+r,\vec{x}+\vec{Y}(t)\big)\,d^{4}x,\label{ABnoRet1}
\end{equation}
where  now $d^{4}x\equiv dtd^{3}x$, $r\equiv\left|\vec{x}\right|$ and,
we recall, ${\mathcal V}^\m(t)=(1,\vec V(t))$. Since the world-line is time-like it is a textbook exercise\footnote{For bounded trajectories the proof is rather straightforward, see {\it e.g.}  Section \ref{ABProof}. For unbounded (asymptotically linear) trajectories one may use that for large negative $t$ one has $\vec Y(t)\approx t\vec V_\infty$, with $|\vec V_\infty|<1$.} to show that the linear functionals \eref{ABnoRet1} represent indeed   distributions, that is that they can be dominated by a finite sum of {\it semi-norms}, see \eref{distDef}.

The Li\'enard-Wiechert {\it field} derived from  $\eqref{AV}$ can be written as the sum
\begin{equation}
F^{\mu\nu}=C^{\mu\nu}+R^{\mu\nu},\label{FVdec}
\end{equation}
where $C^{\mu\nu}$ represents the {\it Coulomb} field
\begin{equation}
C^{\mu\nu}=\frac{e}{4\pi}\frac{(L^{\mu}U^{\nu}-L^{\nu}U^{\mu})U^{2}}
{\left(UL\right)^{3}},\label{CV}
\end{equation}
decreasing as $1/r^2$ at large distances from the particle,
and the second term represents the {\it radiation} field
\begin{equation}
R^{\mu\nu}=\frac{e}{4\pi}\frac{L^{\mu}((UL)W^{\nu}-(WL)
U^{\nu})}{(UL)^{3}}-(\mu\leftrightarrow\nu),\label{RV}
\end{equation}
that at large distances decreases as $1/r$. These fields satisfy the equations
\begin{equation}
\partial_{[\mu}C_{\nu\rho]}=0,\quad\quad\partial_{[\mu}R_{\nu\rho]}=0,\label{CRVBianchi}
\end{equation}
\begin{equation}
\partial_{\mu}C^{\mu\nu}=J^{\nu}+K^{\nu},\quad\quad\partial_{\mu}R^{\mu\nu}=-K^{\nu},
\label{CRVMaxwell}
\end{equation}
where we introduced the vector field
\begin{equation}
K^{\mu}=\frac{e}{2\pi}\frac{(WL)U^2}{(UL)^{4}}\,L^{\mu}.\label{K}
\end{equation}
In these derivations a crucial role is played the identity, implied by $L^2=0$,
\be\label{idl}
\frac{\pa\la(x)}{\pa x^\m}=\frac{L_\m}{(UL)}.
\ee
Notice  that the fields $C^{\mu\nu}$ and  $R^{\mu\nu}$ fulfill  the Bianchi identity  separately, see \eref{CRVBianchi}, while only their sum satisfies the equation $\pa_\m F^{\m\n}=J^\n$.

\subsection{Linear light-like trajectories and failure of the Green function method} \label{GreenFail}

We turn now to our main topic, {\it i.e.} the solution of Maxwell's equations
\begin{equation}
\partial_{\mu}\mathcal{F}^{\mu\nu}=\mathcal{J}^{\nu},\quad\quad
\partial_{[\m}\mathcal{F}_{\n\rho]}=0,\label{MaxEq}
\end{equation}
for a massless charged particle moving along a {\it light-like} world-line $y^{\mu}(\lambda)$ with current
\begin{equation}
\mathcal{J^{\mu}}(x)= e\!\int\!\delta^{4}(x-y(\lambda))\, dy^{\mu}.\label{LLCurrent}
\end{equation}
In analogy to the time-like case we define the four-velocity --  subject now to the light-like condition $u^2=0$ as well as to $u^0>0$ -- through $u^{\mu}={dy^{\mu}}/{d\lambda}$ and the four-acceleration through $w^\m=du^\m/d\la$. We will denote the three-dimensional velocity and acceleration of the particle respectively by $\vec v=d\vec y/dt=\vec u/u^0$ and  $\vec a=d\vec v/dt$ and we will also use the notation $v^{\mu}={dy^{\mu}}/{dt}=\left(1,\vec{v}\right)$.

Solving the Bianchi identity in \eref{MaxEq} in terms of a potential $\mathcal{A}^{\mu}$ leads to ${\cal F}^{\m\n}=\pa^\m {\cal A}^\n-\pa^\n {\cal A}^\m$ and, in the
Lorenz gauge $\partial_{\mu}\mathcal{A}^{\mu}=0$,
equations $\eqref{MaxEq}$ reduce again to the standard form
\begin{equation}
\square\mathcal{A}^{\mu}=\mathcal{J}^{\mu}.\label{boxAURR}
\end{equation}
In principle one could now rely again on the Green function method, based {\it formally} on the equations
\begin{equation}
\mathcal{A}^{\mu}=G*\mathcal{J}^{\mu},\quad\quad G(x)=\frac{1}{2\pi}\,H(x^{0})\,\delta(x^{2}).\label{greenEqq}
\end{equation}
However -- and this is one of the main observations
of the present paper -- if $\mathcal{J}^{\mu}$ is the  current of a massless
particle traveling along a {\it light-like} trajectory, in general the convolution $G*\mathcal{J}^{\mu}$ is not a distribution -- contrary to what happens for time-like trajectories.

We illustrate this feature for a particularly simple, but relevant, trajectory, {\it i.e.} for a {\it linear} motion (which is necessarily uniform) with world-line
\begin{equation}
y^{\mu}(\lambda)=u^{\mu}\lambda,\quad\quad u^{2}=0.\label{URwline}
\end{equation}
Starting from the general expression \eref{AV0} one obtains now, formally,
\begin{alignat}{1}
\mathcal{A^{\mu}}(x)=G*\mathcal{J}^{\mu}
 & =\frac{eu^{\mu}}{2\pi}\int \!H(x^{0}-u^{0}\lambda)\,\delta\!\left((x-u\lambda)^{2}\right) d\lambda\nonumber
=\left.\frac{eu^{\mu}H(ux)}{4\pi\big|(x-u\lambda)^{\alpha}u_{\alpha}\big|}
\right|_{\lambda=x^{2}/2\left(ux\right)}\nn\\
& =\frac{ev^{\mu}}{4\pi} \frac{H(vx)}{(vx)}.\label{aline}
\end{alignat}
Due to the singularity along the line $(vx)= t-\vec v\cdot\vec{x}=0$
this potential is not a locally integrable function and, consequently, it is {\it not} a distribution. {\it A fortiori}  it is {\it not} a solution of equation \eref{boxAURR} -- as is often stated erroneously in the literature. As will become clear in Section \ref{ubm}, the
reason for this pathology is that the trajectory is {\it unbounded} in the
past.

In conclusion, for light-like trajectories in general the Green function method fails to provide a solution to Maxwell's equations $\eqref{MaxEq}$.

\subsection{Potential from a limiting procedure: shock-wave\label{limAUR}}

A general alternative approach to solve Maxwell's equations for light-like
trajectories consists in adopting an appropriate limiting procedure,
starting from the potential of a convenient {\it time-like} trajectory with  {\it constant} regulator speed $V<1$. More precisely, given an arbitrary light-like world-line $y^\m(\la)$ we introduce a time-like {\it regularized}
world-line $Y^{\mu}(\lambda)$ defined by
\begin{equation}
Y^{0}(\lambda)=\frac{y^{0}(\lambda)}{V},\quad\quad \vec{Y}(\lambda)=\vec{y}(\lambda).
\label{YVUR}
\end{equation}
The velocity of this regularized world-line is given by
$$
\vec V(t)=\frac{d\vec Y}{dY^0}=V\vec v\left(Vt\right)
$$
and the regularized motion occurs, thus, at the {\it subluminal} constant speed $|\vec V(t)|=V<1$, along the same orbit of the light-like trajectory.

In the particular case of the linear motion \eref{URwline} we get the regularized world-line $Y^0(\la)=u^0\la/V$, $\vec Y(\la)=\vec u\la$, corresponding to a linear motion with  the constant velocity $\vec V=V\vec v$.
For such a time-like world-line the solution of \eref{boxAUR} is given by the potential \eref{AV}, that for a constant velocity  reduces to the standard result
\begin{equation}
A^{\mu}(x)=\frac{e}{4\pi}\frac{\cal V^{\mu}}{\sqrt{({\cal V}x)^{2}-x^{2}{\cal V}^{2}}},\quad\quad {\cal V}^\m=(1,\vec V).\label{AURV}
\end{equation}
Observe now that
one has the, trivial but important, distributional limit
\begin{equation}
\LimV J^{\mu}=\mathcal{J}^{\mu},\label{LimVJ}
\end{equation}
that, we recall, amounts to the ordinary limits in $\mathbb{C}$
\be\label{lims}
\lim_{V\ra 1} J^{\mu}(\vp)=\mathcal{J}^{\mu}(\vp),
\ee
for every test function $\vp$.
Throughout this paper we denote the {\it distributional} limit with the capital symbol $\mathrm{Lim}_{V\rightarrow1}$,
to distinguish it from the ordinary \textit{point-wise} limit in $\mathbb{C}$, denoted by $\lim_{V\rightarrow1}$.

Due to \eref{LimVJ} one might expect that the distributional limit of the potential \eref{AURV} -- which by construction satisfies $\square A^\m=J^\m$ -- in the limit $V\rightarrow1$ becomes  a solution $\mathcal{A}^{\mu}$
of equation $\eqref{boxAURR}$. This expectation is, however, spoiled by the fact that, while the functions \eref{AURV} admit the {\it point-wise} limits
\be\label{V1}
 \lim_{V\ra 1}A^\m(x)=\frac{e}{4\pi} \frac{v^\m}{|(vx)|},
\ee
they do not admit {\it distributional} limits. At the same footing of \eref{aline} the functions at the r.h.s. of \eref{V1} are, in fact, not distributions\footnote{The difference between \eref{aline} and \eref{V1} is meaningless, since  both are ill-defined as distributions.}.

To overcome this difficulty we take advantage from the fact that $A^{\mu}$ is defined
up to a gauge transformation $
A^{\mu}\rightarrow A^{\mu}+\partial^{\mu}\Lambda.$
Choosing as gauge function
\begin{equation}
\Lambda=\frac{e}{4\pi}\ln\left|({\cal V}x)-\sqrt{({\cal V}x)^{2}-x^{2}{\cal V}^2}\,\right|,
\label{gaugeT}
\end{equation}
 the transformed potential reads
\begin{equation}
\widetilde{A}^{\mu}=A^{\mu}+\partial^{\mu}\Lambda=
\frac{ex^\m}{4\pi}\bigg(\!1+\frac{({\cal V}x)}
{\sqrt{({\cal V}x)^{2}-x^{2}{\cal V}^{2}}}
\bigg)
\mathcal{P}\!\left(\frac{1}{x^{2}}\right)\!,\label{AURVreg}
\end{equation}
where $\mathcal{P}$ stands for the {\it principal part}. We have obviously
\begin{equation}
F^{\mu\nu}=
\partial^{\mu}{A}^{\nu}-\partial^{\nu}{A}^{\mu}=\partial^{\mu}\widetilde{A}^{\nu}
-\partial^{\nu}\widetilde{A}^{\mu},
\label{FVURaltGauge}
\end{equation}
but now the distributional limit
\begin{equation}
\mathcal{A}^{\mu}\equiv\LimV\widetilde{A}^{\mu}=\frac{ex^\m}{2\pi}\,H(vx)\,\mathcal{P}\!
\left(\frac{1}{x^{2}}\right)\!
\label{AUR}
\end{equation}
exists and correspondingly $\mathcal{A}^{\mu}$ is a distribution.
Since in the space of distributions derivatives are continuous operators, the
distributional limit of the second expression in $\eqref{FVURaltGauge}$
exists too and, moreover, we can exchange limits with derivatives:
\begin{equation}
\LimV F^{\mu\nu}=\partial^{\mu}\LimV\widetilde{A}^{\nu}-\partial^{\nu}
\LimV\widetilde{A}^{\mu}=\partial^{\mu}\mathcal{A}^{\nu}-\partial^{\nu}\mathcal{A}^{\mu}
\equiv\mathcal{F}^{\mu\nu}.\label{LimVF}
\end{equation}
Given the limits $\eqref{LimVJ}$ and \eref{LimVF}, applying the distributional
limit for $V\rightarrow1$ to the system $\eqref{MaxEqV}$ we conclude, therefore,
that $\mathcal{F}^{\mu\nu}$  {\it satisfies indeed Maxwell's equations}
$\eqref{MaxEq}$. The procedure we have just outlined -- appropriately
generalized -- will be applied throughout this paper to derive exact solutions
of Maxwell's equations for more general trajectories.

Computing the curl of the potential \eref{AUR} one obtains eventually the known electromagnetic field of a {\it shock-wave} -- proportional to a $\dl$-function --
\begin{equation}
\mathcal{F}^{\mu\nu}=\frac{e}{2\pi}\frac{v^{\mu}x^{\nu}-v^{\nu}x^{\mu}}{x^{2}}\,\delta
(vx),\label{FUR}
\end{equation}
vanishing everywhere, except on a plane perpendicular to the trajectory and passing through the particle's position. The plane moves thus together with the particle at the speed of light.

Notice eventually that the potential $\eqref{AUR}$ does {\it not}
satisfy the Lorenz gauge. This gauge can however be restored through a further
gauge transformation, in that from \eref{AUR} one obtains (see also \cite{B})
\be\label{lorg}
{\mathcal{A}}^{'\!\mu}\equiv
\mathcal{A}^{\mu}-\partial^{\mu}\!\left(\frac{e}{4\pi}H(vx)\ln|x^{2}|\right)
=-\frac{ev^\m}{4\pi}\ln|x^{2}|\,
\delta(vx),\quad\quad\partial_{\mu}
{\mathcal{A}}^{'\!\mu}=0.
\ee
Notice  that ${\mathcal{A}}^{'\!\mu}$ looks rather different from the (wrong) Lorenz-gauge potential \eref{aline} -- derived with the Green function method -- although both are proportional to the constant vector $v^\m$.

\subsection{The field from a distributional limit \label{secDistLimR}}

An alternative way to solve Maxwell's equations \eref{MaxEq} consists
in applying the distributional limit directly to the system \eref{MaxEqV},
that is solved by the field strength $F^{\m\n}=\partial^{\mu}A^{\nu}-\partial^{\nu}A^{\mu}$ of the regularized potential \eref{AURV}, that is
\begin{equation}
F^{\mu\nu}=\frac{e}{4\pi}\frac{(x^{\mu}{\cal V}^{\nu}-x^{\nu}{\cal V}^{\mu}){\cal V}^{2}}
{\left(({\cal V}x)^{2}-x^{2}{\cal V}^{2}\right)^{3/2}}.\label{FURV}
\end{equation}
If this field  admits a distributional limit
\be\label{fomn}
\mathcal{F}^{\mu\nu}\equiv \LimV F^{\mu\nu},
\ee
then, thanks to \eref{LimVJ} and to the continuity of derivatives in the space of distributions, the field $\mathcal{F}^{\m\n}$ satisfies automatically  Maxwell's equations \eref{MaxEq}.

To prove \eref{fomn} we must establish the existence of the ordinary limits
\begin{equation}
\mathcal{F}^{\mu\nu}(\varphi)= \lim_{V\rightarrow1}F^{\mu\nu}(\varphi),\quad\quad\forall \,\varphi\in\mathcal{S}.
\label{limFV}
\end{equation}
Without loss of generality we may set $\vec{v}=\left(0,0,1\right)$ so that ${\cal V}^\m=(1,0,0,V)$, see  \eref{AURV}. From \eref{FURV}
we get then
\begin{alignat*}{1}
F^{\mu\nu}(\varphi) & =\int \!F^{\mu\nu}(x)\,\varphi(x)\,d^{4}x
=\frac{(1-V^{2})\,e}{4\pi}\int\!\frac{x^{\mu}\mathcal{V}^{\nu}-x^{\nu}
\mathcal{V}^{\mu}}{\big((\mathcal{V}x)^{2}-(1-V^{2})x^{2}
\big)^{3/2}}\,\varphi(x)\, d^{4}x\\
 &=\frac{(1-V^{2})\,e}{4\pi}\int\!\frac{x^{\mu}\mathcal{V}^{\nu}-x^{\nu}
\mathcal{V}^{\mu}}{\big((z-Vt)^{2}-(1-V^{2})
(x^{2}+y^{2})\big)^{3/2}}\,\varphi(x)\, d^{4}x.
\end{alignat*}
Performing the shift $z\rightarrow z+Vt$ and rescaling
 $z\rightarrow\sqrt{1-V^{2}}z$
this integral becomes
\begin{equation}
F^{\mu\nu}(\varphi)=\frac{e}{4\pi}\int\!\frac{\xi{}^{\mu}\mathcal{V}^{\nu}-\xi^{\nu}
\mathcal{V}^{\mu}}{\left(z^{2}+x^{2}+y^{2}\right)^{3/2}}\,
\varphi(t,x,y,\sqrt{1-V^{2}}z+Vt)\, d^{4}x,\label{limURFmid}
\end{equation}
where  $\xi^\m=(t,x,y,\sqrt{1-V^{2}}z+Vt)$.
Resorting to the {\it dominated convergence theorem}\footnote{To apply this theorem
one must show that the integrand in \eref{limURFmid} can be uniformly (in $V$) dominated by an integrable function. This procedure is standard, although sometimes a bit cumbersome, and consequently throughout the paper we will usually apply it without furnishing the details. For the physically meaningful case
of the potential of an unbounded motion, however, we present the details in Appendix V.}, see Appendix I,  we can now take the limit $V\ra1$ under the integral sign -- in this limit we have in particular $\mathcal{V}^{\mu}\ra v^{\mu}$ and $\xi^\m\ra (t,x,y,t)$ -- so that the integral over $z$ becomes trivial.
The result can be written in the form
\[
\lim_{V\rightarrow1}F^{\mu\nu}(\varphi) =\frac{e}{2\pi}\int\frac{\left(v{}^{\mu}x^{\nu}-v{}^{\nu}x^{\mu}\right)}{x^\rho x_\rho}\,
\delta(t-z)\,\varphi(x)\, d^{4}x,
\]
that matches with $\eqref{FUR}$.

\section{Bounded motion\label{bm}}

In this section we solve equations $\eqref{MaxEq}$ for a generic time-like
{\it bounded} motion, that is a trajectory for which $\left|\vec y(t)\right|<M$, $\forall \,t$. As we will see, in this case the potential
$\mathcal{A}^{\mu}$ derived through the Green function method defines
indeed a distribution. The resulting field is then given by
\begin{equation}
\mathcal{F}^{\mu\nu}=\partial^{\mu}\mathcal{A}^{\nu}-\partial^{\nu}\mathcal{A}^{\mu},
\label{FbFromA}
\end{equation}
where -- we insist -- the derivatives must be computed in the sense of distributions.
Although a distribution admits always partial derivatives, in the presence of singularities, as the ones entailed by $\mathcal{A}^\m$, see below, their explicit evaluation may not be that straightforward.
To compute them we shall adopt a regularization procedure -- playing a role similar to the Li\'enard-Wiechert regularization employed in Sections \ref{limAUR} and \ref{secDistLimR} -- replacing $\mathcal{A^{\mu}}$ with a regularized potential $\mathcal{A_{\varepsilon}^{\mu}}$ of class $C^\infty$ ($\ve>0$ being a constant regulator with the dimension of length) that under the {\it distributional} limit $\ve\ra 0$ tends to $\mathcal{A}^{\mu}$.
Since the potential
$\mathcal{A_{\varepsilon}^{\mu}}$ is of class $C^\infty$, we will be allowed to evaluate its curl $\mathcal{F}^{\mu\nu}_\ve\equiv \partial^{\mu}\mathcal{A}^{\nu}_\ve-\partial^{\nu}\mathcal{A}^{\mu}_\ve$
in the sense of \textit{functions}. The actual field $\eqref{FbFromA}$ can then be recovered as the distributional limit of the field $\mathcal{F}^{\mu\nu}_\ve$ as  $\ve\ra 0$.

Contrary to the shock-wave field $\eqref{FUR}$ of a linear (unbounded) motion, the  field $\eqref{FbFromA}$ will show up $\dl$-like singularities supported on a manifold of lower dimensions:
the {\it planar} shock-wave will in fact be replaced by a  $\dl$-function supported on
a {\it string} attached to the particle.

As anticipated in the introduction, even if the Green function method furnishes a potential $\mathcal{A}^\m$ that is a well-defined distribution, due to the singularities involved it is by no means guaranteed that it satisfies Maxwell's equations: the computation
$$
\square{ \cal A}^\m=\square(G*{\cal J^\m})=(\square G)* {\cal J}^\m=\dl^4* \cal J^\m=\cal J^\m
$$
has indeed only {\it formal} validity, when the distributions $G$ and/or ${\mathcal J}$ are too singular. This means that in general one must prove {\it a posteriori} that the potential $\mathcal{A}^\m$ derived with the Green function method solves Maxwell's equations.

\subsection{Singularities of light-like trajectories}\label{singl}

The Li\'enard-Wiechert fields \eref{CV} and \eref{RV} are singular {\it only} on the world-line itself, {\it i.e.} if $x^\m=Y^\m(\la_0)$ for some $\la_0$. Although the expressions \eref{CV} and \eref{RV} are valid only for time-like world-lines, we can use them to infer the locations of the singularities of the field generated by particles traveling along light-like trajectories.

As long as $U^{\mu}$ is time-like, the denominators $(UL)^{3}$ in
$\eqref{CV}$ and $\eqref{RV}$  vanish only on the trajectory, since the contraction between a non-vanishing light-like and a time-like vector never vanishes. This is no
longer true when the particle moves along a
light-like world-line $y^{\mu}(\lambda)$, with four-velocity $u^{\mu}(\lambda)=dy^{\mu}/d\lambda$ satisfying $u^{2}=0$. In this case the quantity $(UL)$ is replaced by $(ul)$, where
\[
l^\m=x^\m-y^\m(\la),
\]
and now the scalar product $(ul)=u^0\,l^0-\vec u\cdot\vec l =|\vec u|\,|\vec l|-\vec u\cdot\vec l$  vanishes, indeed, also when the three-vectors $\vec{u}$ and $\vec{l}=\vec x-\vec y(\la)$ are {\it aligned}, {\it i.e.}  when
\be\label{alig}
\vec v(\la)=\frac{\vec u(\la)}{|\vec u(\la)|}=\frac{\vec{x}-\vec{y}(\lambda)}{
|\vec{x}-\vec{y}(\lambda)|}.
\ee
Parameterizing the  word-line with time, that is setting $y^\m(\la)=(\la,\vec y(\la))$, the  relations  \eref{lret} amount now to
\[
\left|\vec{x}-\vec{y}(\lambda)\right|=t-\lambda,
\]
so that  \eref{alig} furnishes the singularity locus
\[
\vec{x}=\vec{y}(\lambda)+(t-\lambda)\vec{v}(\lambda), \quad \quad t-\la\ge0.\label{prgamma}
\]
Setting  $b=t-\lambda$ we obtain the {\it singularity-string} at time $t$
\begin{equation}
\vec{\gamma}\left(t,b\right)\equiv\vec{y}(t-b)+b\vec{v}(t-b), \quad\quad b\ge0,\label{gamma}
\end{equation}
along which the field of a massless particle is thus expected to be singular.
Due to the condition $b\ge 0$ the string has one endpoint at the particle's position,  $\vec{\gamma}(t,0)=\vec{y}(t)$,  and
as the particle moves the string $\eqref{gamma}$ sweeps
out a  two-dimensional {\it singularity-surface}, that after a rescaling of $b$ can be rewritten in the covariant form
\begin{equation}
\Gamma^{\mu}(\lambda,b)=y^{\mu}(\lambda)+bu^{\mu}(\lambda), \quad\quad b\ge0.\label{Gamma}
\end{equation}
\textbf{Poincar\'e-duality.} In the following a special role will be
played by a particular distribution  proportional to a $\delta$-function
supported on the surface $\Gamma^\m$. In the same way as one associates
to a world-line $y^{\mu}(\lambda)$ the distribution-valued vector
field $\mathcal{J}^{\mu}$ $\eqref{LLCurrent}$, to an arbitrary regular
surface $\Gamma^{\mu}(\lambda,b)$ one can
associate the antisymmetric reparameterization invariant (distribution-valued)  field
\begin{equation}
Q^{\mu\nu}(x)=e\int\frac{\partial\Gamma^{\mu}(\lambda,b)}
{\partial b}\frac{\partial\Gamma^{\nu}(\lambda,b)}{\partial\lambda}\, \delta^4(x-\Gamma(\lambda,b)) \,db\,d\lambda-(\mu\leftrightarrow\nu).\label{pmn}
\end{equation}
The corresponding map, that in general associates to a $p$-submanifold
an antisymmetric tensor field of rank $p$ supported on that submanifold,
goes under the name of \textit{Poincar\'e-duality}\footnote{Actually in the framework of Differential Geometry Poincar\'e-duality associates
to a $p$-submanifold in $D$ dimensions a $(D-p)$-form,
that is the Hodge dual of the rank-$p$ tensor \eref{pmn},
with $e=1$.}. In the particular case of the surface $\eqref{Gamma}$ the field $\eqref{pmn}$
becomes
\begin{equation}
Q^{\mu\nu}(x)=e\!\int\! bH(b)\big(u^{\mu}(\la)w^{\nu}(\la)-u^{\nu}(\la)w^{\mu}(\la)\big)\,
\delta^{4}\big(x-\Gamma(\lambda,b)\big)db
d\lambda,\label{P}
\end{equation}
that applied  to a test function gives
\begin{alignat}{1}
Q^{i0}(\vp)&=-e\int_0^\infty\!b db\int_{-\infty}^\infty dt\,a^i\vp(t+b,\vec y+b\vec v),\label{pi0}\\
Q^{ij}(\vp)&=e\int_0^\infty \!bdb\int_{-\infty}^\infty dt\,(v^ia^j-v^ja^i)\,\vp(t+b,\vec y+b\vec v),\label{pij}
\end{alignat}
where the variables $\vec y$, $\vec v$ and $\vec a$ are evaluated at $t$.

A particular feature of Poincar\'e-duality is that it associates to
the \textit{boundary} of a manifold the \textit{divergence} of
the field associated to the manifold. If the trajectory $\vec y(t)$ is {\it bounded}, the boundary of the surface $\Gamma^\m$ in $\eqref{Gamma}$ is precisely the world-line $y^{\mu}(\lambda)$. In fact, in this case from \eref{gamma} one sees that $\vec\gamma(t,0)=\vec y(t)$, while under $b\ra \infty$  $\vec \gamma(t,b)$ tends to a point at infinity. We have thus
\begin{equation}
\partial_{\mu}Q^{\mu\nu}=\mathcal{J}^{\nu},\label{dpmn}
\end{equation}
an identity that will prove to be useful later. If, on the contrary, the trajectory is {\it unbounded}, the boundary of the surface $\Gamma^\m$ may acquire an additional component and equation \eref{dpmn} will then be modified, see equation \eref{upmn}.

\subsection{A distribution-valued potential\label{ABProof}}

For a generic light-like trajectory the Green function method  -- see \eref{greenEqq} and \eref{LLCurrent} -- formally gives the potential
\be\label{amuf}
\mathcal{A}^{\mu}=G*\mathcal{J}^{\mu}=\frac{e}{4\pi}\left.\frac{u^{\mu}}{\left(ul\right)}
\right|_{\lambda=\lambda(x)},
\ee
where we proceeded as in \eref{convAUR}-\eref{AV}.
This time $l^\m(x)\equiv x^\m-y^\m(\la(x))$ and $\la(x)$ is determined by the conditions
\be\label{llcaus}
(x-y(\la))^2=0,\quad\quad x^0>y^0(\la).
\ee
To check whether or not the four functions \eref{amuf} are distributions we must apply them to a test function $\vp$. With computations analogous to those that led from \eref{ABconv} to \eref{ABnoRet1} we obtain
\begin{equation}
\mathcal{A^{\mu}}(\varphi)=\frac{e}{4\pi}\!\int\frac{v^{\mu}(t)}{r}\,
\varphi(t+r,\vec{x}+\vec{y}(t))\,d^{4}x,\quad \quad v^\m(t)=(1,\vec v(t)),\quad r=|\vec x|.\label{ABnoRet}
\end{equation}
We can  now show  that these integrals
obey the following theorem.
\vskip0.2truecm\noindent
{\bf Theorem.} {\it The four linear functionals on ${\cal S}(\mathbb{R}^{4})$ given in \eref{ABnoRet} represent distributions}.
\vskip0.2truecm
\noindent {\it Proof}. By definition we must prove that the integrals
$\eqref{ABnoRet}$ satisfy the bound $\eqref{distDef}$ in terms
of semi-norms. Since  for each $\m$ we have $|v^{\mu}(t)|\leq1\;\forall \,t$, performing  the shift $t\rightarrow t-r$ we get the estimate
\begin{alignat*}{1}
\left|\mathcal{A^{\mu}}(\varphi)\right| & \leq\frac{e}{4\pi}\int\frac{|\,\varphi(t,\vec{x}+\vec{y})|}{r}\,d^{4}x\\
 & =\frac{e}{4\pi}\int\frac{(1+t^{2})\big(1+|\vec{x}+\vec{y}|^{2}
\big)^{2}|\,\varphi(t,\vec{x}+\vec{y})|}{r(1+t^{2})\big(1+|\vec{x}+\vec{y}|^{2}\big)^{2}}\,d^{4}x,
\end{alignat*}
where from now on $\vec{y}\equiv\vec{y}(t-r)$. For a {\it bounded} motion we have moreover the uniform estimate
\begin{equation}
\left|\vec{x}+\vec{y}\right|\geq\left|\vec{x}\right|-\left|\vec{y}\right|\geq r-M,\label{ABDistProof1}
\end{equation}
so that
\be\label{aint}
\left|\mathcal{A^{\mu}}(\varphi)\right|\leq\frac{e\Vert\vp \Vert}{4\pi}\int\!\frac{1}
{r(1+t^{2})\big(1+(r-M)^{2}H(r-M)\big)^{2}}\,d^{4}x,
\ee
where
\[
\Vert\vp \Vert \equiv \sup_{x\in\mathbb{R}^{4}}\left((1+t^{2})
\big(1+|\vec{x}+\vec{y}|^{2}\big)^{2}|\,\varphi(t,\vec{x}+\vec{y})|\right)
\]
is a finite linear combination of semi-norms of $\varphi$ and the (converging) integral in \eref{aint} is a positive constant. The bound \eref{aint} is therefore of the form \eref{distDef}.  \hfill $\square$
\\

For an {\it unbounded} light-like motion $\eqref{ABDistProof1}$ is no
longer valid. Even worse, in that case the integral $\eqref{ABnoRet}$
is in general  \textit{divergent}. Consider, for example, a trajectory
that for $t\rightarrow-\infty$ becomes asymptotically linear: $\vec{y}(t)\approx\vec{v}_{\infty}t$,
where $\vec{v}_{\infty}$ is a constant velocity with $\left|\vec{v}_{\infty}\right|=1$.
For such a  world-line the test function in $\eqref{ABnoRet}$ for large negative values of $t$ behaves  as $\varphi(t+r,\vec{x}+\vec{v}_{\infty}t)$ and along the line $\vec{x}=-\vec{v}_{\infty}t$,
for which $r=\left|\vec{x}\right|=-t$, in general it
does not vanish for any $r$ -- in that its arguments remain always finite --
while the rest of the integrand, in polar coordinates, grows as $r$, leading to a divergence. In particular for a linear trajectory we retrieve that the potential \eref{aline} is not a distribution.

\subsection{Derivation of the electromagnetic field}

\subsubsection{Covariant regularization}\label{covreg}

As we pointed out previously (Section \ref{singl}), on the surface $\Gamma$ $\eqref{Gamma}$ the potential \eref{amuf} is singular and correspondingly the evaluation of the field strength
$
\mathcal{F}^{\mu\nu}=\partial^{\mu}\mathcal{A}^{\nu}-\partial^{\nu}\mathcal{A}^{\mu}
$
requires to compute the derivatives in the sense
of distributions. As anticipated in the introduction of this section, to this order it is helpful to resort to a regularized potential
$\mathcal{A}_{\varepsilon}^{\mu}$ of class  $C^\infty$.

Before presenting our regularization we notice, however, that in the complement  of $\Gamma$ -- where the potential \eref{amuf} is of class $C^\infty$ -- its derivatives can actually be computed in the ordinary sense of functions. With a standard calculation, based on the relation $\pa_\m\la(x)=l_\m/(ul)$ following from  \eref{llcaus}, we find the expected result
\be
\big(\partial^{\mu}\mathcal{A}^{\nu}-\partial^{\nu}\mathcal{A}^{\mu}\big)\big|_{{\mathbb R}^4\backslash \Gamma}=\frac{e}{4\pi}\frac{l^{\mu}((ul)w^{\nu}
-(wl)u^{\nu})}{(ul)^{3}}
-(\mu\leftrightarrow\nu)\equiv
\mathcal{F}^{\mu\nu}_{reg}.
\label{RBpointWise}
\ee
This means that the actual field  $\mathcal{F}^{\mu\nu}$ can differ from the ``regular'' distribution $\mathcal{F}^{\mu\nu}_{reg}$ only through terms supported on $\Gamma$. The presence (or absence) of such terms can, however,  be revealed only through a distributional calculation.

A convenient regularization is obtained replacing the Green function $G(x)$ of the d'Alembertian by
\begin{equation}
G_{\varepsilon}(x)=\frac{1}{2\pi}\,H(x^{0})\,\delta(x^{2}-\varepsilon^{2}),\label{Geps}
\end{equation}
where $\ve>0$ is a regulator with the dimension of length.
Proceeding  as in \eref{convAUR}-\eref{AV} one obtains then the $C^\infty$-potential
\be\label{ame}
\mathcal{A_{\varepsilon}^{\mu}}= G_\ve*{\cal J}^\m =\frac{e}{4\pi}\left.\frac{u^{\mu}}{\left(ul\right)}
\right|_{\lambda=\lambda_{\varepsilon}(x)},
\ee
where  the function $\lambda_{\varepsilon}(x)$ is determined by the conditions
\begin{equation}
l^2=\left(x-y(\lambda_{\varepsilon})\right)^{2}=\varepsilon^{2},\quad \quad x^{0}\geq y^{0}(\lambda_{\varepsilon}),\label{lretEps}
\end{equation}
replacing \eref{llcaus}. Notice that, since $l^\m\equiv x^\m-y^\m(\la_\ve(x))$ is  time-like and $u^\m$ light-like, the denominator $(ul)=u^\m l_\m$ {\it never} vanishes, not even on the world-line: this ensures that the potential \eref{ame} is indeed of class $C^\infty$.

Employing calculations similar to those performed in \eref{ABconv}-\eref{ABnoRet1}, from \eref{ame} we obtain the regularized functionals, obviously defining distributions,
 \begin{equation}
\mathcal{A_{\varepsilon}^{\mu}}(\varphi)=\frac{e}{4\pi}\int\frac{v^{\mu}(t)}
{r_{\varepsilon}}\,\varphi(t+r_{\varepsilon},\vec{x}+\vec{y}(t))\,d^{4}x,\label{ABepsNoret}
\end{equation}
which differ from \eref{ABnoRet} only through the replacement
$r\ra  r_{\varepsilon}=\sqrt{r^{2}+\varepsilon^{2}}$. Correspondingly it
is straightforward to show that one has the distributional limits
\begin{equation}
\LimE\mathcal{A}_{\varepsilon}^{\mu}=\mathcal{A}^{\mu},\label{LimABeps}
\end{equation}
so that the potential \eref{ame} represents indeed a regularization of the potential \eref{amuf}. From \eref{ame} it follows in particular that
this regularization has the advantage of being manifestly \textit{Lorenz-covariant}. On the other hand, as we will see in Section \ref{LWreg},  it is suitable only for bounded motions, and consequently for unbounded trajectories we have to resort to the alternative Li\'enard-Wiechert-regularization \eref{YVUR}.

The distributional derivatives of \eref{ame} showing up in the regularized Maxwell field \be\label{freg}
\mathcal{F}^{\mu\nu}_\ve=\partial^{\mu}\mathcal{A}^{\nu}_\ve
-\partial^{\nu}\mathcal{A}^{\mu}_\ve
\ee
can now be computed as ordinary derivatives and accordingly the result can be retrieved from $\eqref{FVdec}$-\eref{RV}. However,
since in this case we have $u^{2}=0$, the regularized Coulomb field \eref{CV} \textit{vanishes} --
and of course also its distributional limit is zero!
We are therefore left only with
the regularized {\it radiation} field
\begin{equation}
\mathcal{F}_{\varepsilon}^{\mu\nu}=\frac{e}{4\pi}\left.\frac{l^{\mu}((ul)w^{\nu}
-(wl)u^{\nu})}{(ul)^{3}}\right|_{\lambda=\lambda_{\varepsilon}(x)}
-(\mu\leftrightarrow\nu).\label{FBeps}
\end{equation}
It is also straightforward to check that the potential \eref{ame} preserves the Lorenz-gauge
$
\pa_\m {\cal A}^\m_\ve=0.
$

Eventually we {\it define} the regularized current as
\begin{equation}
\mathcal{J}_{\varepsilon}^{\nu}\equiv \partial_{\mu}\mathcal{F}_{\varepsilon}^{\mu\nu}.\label{Jeps}
\end{equation}
A direct, although a bit lengthy, calculation from \eref{FBeps} furnishes the explicit expression (see \cite{LPAM})
\begin{equation}
\mathcal{J}_{\varepsilon}^{\mu}=\frac{\varepsilon^{2}e}{4\pi}\left(\!\frac{
b^\m(ul)-u^{\mu}(bl)}{(ul)^{4}}
+\frac{3(wl)}{(ul)^{5}}\big((wl)u^{\mu}-(ul)
w^{\mu}\big)\!\right)\!, \quad b^\m\equiv \frac{dw^\m}{d\la}.  \label{Jreg}
\end{equation}
Notice that  this current is proportional to  $\ve^2$, implying that its limit under $\ve\ra 0$ is supported necessarily on (a subset of) the singularity surface \eref{Gamma}.

\subsubsection{Solving Maxwell's equations}

Equations $\eqref{LimABeps}$ and \eref{freg} imply that the distributional limit of the field \eref{FBeps} under $\ve\ra 0$ exists and, moreover, that
\begin{equation}
\mathcal{F}^{\mu\nu}\equiv\partial^{\mu}\mathcal{A}^{\nu}-\partial^{\nu}\mathcal{A}^{\mu}
= \LimE\mathcal{F}_{\varepsilon}^{\mu\nu}.\label{FBfromLim}
\end{equation}
Consequently from the definition \eref{Jeps} it follows that also the limit $\LimE\mathcal{J}_{\varepsilon}^{\mu}$ exists, and in Appendix II we prove that moreover \begin{equation}
\LimE\mathcal{J}_{\varepsilon}^{\mu}=\mathcal{J}^{\mu},\label{LimJeps}
\end{equation}
as one might expect\footnote{As a short-cut we note the following: from \eref{Jeps} it follows that the vector field $\LimE\mathcal{J}_{\varepsilon}^{\mu}$ has vanishing divergence and, as observed in the text, its support must belong to the singularity surface \eref{Gamma}. Thanks to Lorentz-invariance it must then be proportional to $\mathcal{J}^{\mu}$.}. Applying the distributional limit $\ve\ra 0$ to equation \eref{Jeps} we conclude then that $\mathcal{F}^{\mu\nu}$ satisfies the first Maxwell equation in \eref{MaxEq}, and from \eref{FBfromLim} it follows that it satisfies also the second Maxwell equation, {\it i.e.} the Bianchi identity. It remains now to determine $\mathcal{F}^{\mu\nu}$  explicitly.

\subsubsection{Determination of the field\label{lof}}

To evaluate the electromagnetic field we have to perform the limit $\eqref{FBfromLim}$ explicitly, {\it i.e.} for every test function we must compute the ordinary limits
\[
\mathcal{F}^{\mu\nu}(\varphi)=\lim_{\varepsilon\rightarrow0}\mathcal{F}_{\varepsilon}^{\mu\nu}
(\varphi),
\]
with $\mathcal{F}_{\varepsilon}^{\mu\nu}(x)$ given in \eref{FBeps}. As this function is of the form $f(x,\la_\ve(x))$ we consider the identity ($l^\m\equiv x^\m-y^\m(\la_\varepsilon)$)
\be\label{fxe}
f(x,\la_\ve(x))=\int\!\dl(\la-\la_\ve(x))f(x,\la)\,d\la
=2\!\int\!\! H(l^0)\,\dl\!\left(l^2-\ve^2\right)(u^\m(\la)l_\m)f(x,\la)\,d\la.
\ee
When applying such a function,  as \eref{FBeps}, to a test function $\vp(x)$ we may  successively perform the shift $x\ra x+y(\la)$ getting
\begin{alignat}{1}\nn
\mathcal{F}_{\varepsilon}^{\mu\nu}(\varphi)&=\frac{e}{2\pi} \!\int\! \! H(x^0)\,\dl(x^2-\ve^2)\,\frac{x^{\mu}((ux)w^{\nu}-(wx)u^{\nu})}{(ux)^{2}}\,\vp(x+y) \,d^4x\,d\la -(\mu\leftrightarrow\nu)\\
&=\frac{e}{4\pi} \!\int \frac{X^{\mu}((uX)w^{\nu}-(wX)u^{\nu})}{r_\ve(uX)^{2}}\,\vp(X+y) \,d^3x\,d\la -(\mu\leftrightarrow\nu),\label{feps}
\end{alignat}
where we have set $X^\m=(r_\ve,\vec x)$ and
\[
r_\ve=\sqrt{r^2+\ve^2}.\]
For definiteness we evaluate  now the limit ${\mathcal F}^{i0}=
\LimE {\cal F}^{i0}_\ve$ regarding the electric field, the procedure for the magnetic field ${\cal F}^{ij}_\ve$ being completely analogous. Taking advantage from reparametrization invariance to choose $\la=y^0(\la)\equiv t$ \eref{feps} gives
\begin{equation}
{\cal F}^{i0}_\ve(\varphi)=\frac{e}{4\pi}\!\int\frac{
(x^i-r_{\varepsilon}v^{i})\,\vec{a}\cdot\vec{x}-r_{\varepsilon}(r_{\varepsilon}
-\vec{v}\cdot\vec{x})\,a^{i}}
{r_{\varepsilon}(r_{\varepsilon}-\vec{x}\cdot\vec{v})^{2}}\,
\varphi(t+r_{\varepsilon},\vec{x}+\vec{y})\,d^3x\,dt,\label{EBProof1}
\end{equation}
where the kinematical variables $\vec y$, $\vec v$ and $\vec a$ are now evaluated at $t$.
As $\varepsilon$ approaches $0$, the denominator
in $\eqref{EBProof1}$ vanishes along the half-line $\vec{x}=b\vec{v}$,  $b>0$,
this half-line being precisely the image in these coordinates at fixed $t$ of the string  $\eqref{gamma}$, where the field becomes indeed singular. To isolate this string
we change coordinates from $\vec x$ to  $(b,q_a)$, $a=1,2$, according to
\begin{equation}
\vec{x}=b\vec{v}+q_{a}\vec{N}_{a},\label{bqVar}
\end{equation}
where we introduced the orthonormal basis $\{\vec{v},\vec{N}_a\}$ at fixed time:
\be\label{bq}
\vec{N}_{a}\cdot\vec{N}_{b}=\delta_{ab},\quad \vec v\cdot\vec v=1,\quad   \vec{N}_{a}\cdot\vec{v}=0, \quad N^i_aN^j_a+v^iv^j=\delta^{ij}.
\ee
In this way the position of the singularity string amounts now to $q_a=0$.

We evaluate the limit of \eref{EBProof1} explicitly in Appendix III the  result being
\begin{equation}
{\cal F}^{i0}(\varphi)=
\lim_{\ve\ra0}{\cal F}^{i0}_\ve(\varphi)
=\frac{1}{2}\,Q^{i0}(\varphi)+\!\int\! H^{i}(b,q,t)\,\varphi(t+r,b\vec{v}+q_{a}\vec{N}_{a}+\vec{y})\,dtdbd^{2}q, \label{EBProof5}
\end{equation}
where
\be\label{hi}
H^{i}(b,{q},t)=\frac{e}{4\pi}\left(\frac{\Pi_{ab}N_{a}^{j}N_{b}^{i}
(r+b)^{2}a^j}{r(q^{2})^{2}}
-\frac{a^{i}}{2r}-\frac{\big(\vec{a}\cdot q_{c}\vec{N}_{c}\big)(r+b)v^i}{rq^{2}}\right)
\ee
and
\[
\Pi_{ab}=q_{a}q_{b}-\frac{q^{2}}{2}\,\delta_{ab},\quad \quad r=\sqrt{b^2+q^2},\quad\quad q=\sqrt{q_1^2+q_2^2}.
\]
The integral in $\eqref{EBProof5}$ is \textit{conditionally convergent},
in the sense that -- by definition -- one must first integrate over the polar angle of $q_a$ and then over its radius  $q$. In this way the  traceless matrix $\Pi_{ab}$ guarantees the convergence of the integral over $q_a$ around $q_a=0$, {\it i.e.} around the singularity-string.

The distribution $Q^{i0}$ in \eref{EBProof5} is the $\dl$-function \eref{pi0} supported on the singularity surface $\Gamma$.
The second term in \eref{EBProof5} arises from the {\it point-wise} limit of ${\cal F}^{i0}_\ve$ and represents thus -- by construction -- the regular field
$\mathcal{F}^{i0}_{reg}$ \eref{RBpointWise}. According to our integration prescription we  indicate  then this term as the ``principal part''
$
\mathcal{P}(\mathcal{F}^{i0}_{reg})(\vp).
$
Thanks to the manifest Lorentz-invariance of our procedure -- that permits us to avoid  the analogous computation for the magnetic components ${\mathcal F}^{ij}$ -- we can then write our final result in the form
\begin{equation}
\mathcal{F^{\mu\nu}}=\frac{1}{2}\,Q^{\mu\nu}+\mathcal{P}(\mathcal{F}^{\mu\nu}_{reg}).\label{FB}
\end{equation}

\subsubsection{Properties of the field}

Analyzing the meaning of the field \eref{FB} we see first of all that there is, indeed, a non-vanishing contribution supported on $\Gamma$, whose structure $Q^{\m\n}$ is essentially determined by Lorentz-invariance. The interpretation of the factor $1/2$ is, instead, less obvious. From $\pa_\m\mathcal{F^{\mu\nu}}={\cal J}^\n=\pa_\m Q^{\mu\nu}$ we  derive
\begin{align}\label{demo1}
\pa_\m\!\left(\frac12\,Q^{\m\n}\!\right)&=\frac12\,\mathcal{J}^\n,\\
\pa_\m\big(\mathcal{P}(\mathcal{F}^{\mu\nu}_{reg})\big) &=\frac12\,\mathcal{J}^\n.\label{demo2}
\end{align}
This means that the flux through a closed surface of the electric field $E^i={\cal F}^{i0}$, {\it i.e.}
\be\label{defe}
\vec E= \vec E_{sing}+\vec{ E}_{reg},\quad\quad
{E}^i_{sing}\equiv \frac12\,Q^{i0},\quad\quad {E}^i_{reg} \equiv \mathcal{P}(\mathcal{F}^{i0}_{reg}),
\ee
is equally distributed between the field $\vec{ E}_{sing}$, supported on the singularity-string, and the regular field
$\vec{E}_{reg}$:
\begin{align}
\vec\nabla\cdot\vec { E}_{sing}&=\frac12\,\mathcal{J}^0, \label{dive0}\\
 \vec\nabla\cdot\vec { E}_{reg}&=\frac12\,\mathcal{J}^0.\label{dive}
\end{align}
While relation \eref{dive0}, we repeat, follows directly from \eref{pi0} and Poincar\'e-duality, the relation \eref{dive} is less obvious in that above we gave an indirect derivation. Correspondingly we provide a direct, and independent, proof of this rather unexpected result in Appendix VIII.

Actually the {\it flux} through a closed surface $S$ of a field like $\vec{ E}_{reg}$, being a distribution, can not be defined through a simple surface integral like $\int_S { E}_{reg}^i\,d\Sigma^i$, if $S$ intersects the singularity curve\footnote{The same is obviously also true for $\vec{ E}_{sing}$, but in that case there is a ``natural'' definition of $\int_S { E}_{sing}^i\,d\Sigma^i$, provided by Poincar\'e-duality, that is in agreement with the formal rules used by physicists to integrate Dirac-$\delta$ distributions.}. In this case to {\it define} the flux one has to proceed as follows. Denote the {\it characteristic function} of the volume $V$ with boundary  $S$ by $\chi(\vec x)$ and introduce a {\it smooth} deformation  $\chi_\gamma(\vec x)\in {\mathcal S}(\mathbb R^3)$ -- vanishing away from  $V$ and equal to 1 well inside $V$ -- that for $\gamma\ra 0$  tends point-wise  to  $\chi(\vec x)$. Using that the distribution
${ E}^i_{reg}$, {\it i.e.} the second term in \eref{EBProof5}, at fixed time defines a distribution in ${\mathcal S}'(\mathbb R^3)$, we can introduce the {\it formal} test function $\varphi_\gamma(t,\vec x)=\delta(t-t_0)\chi_\gamma(\vec x)$
and {\it define} the flux of $\vec{ E}_{reg}$ through $S$ at time $t_0$ as\footnote{It is not necessary to regularize the $\delta$-function $\delta(t-t_0)$, since it can be seen that the distributions $\mathcal{P}(\mathcal{F}^{\mu\nu}_{reg})$ -- belonging to  ${\mathcal S}'({\mathbb R}^4)$ -- at fixed time define elements of ${\mathcal S}'({\mathbb R}^3)$.}
\begin{align}\label{flu}
\Phi_S(t_0)=\lim_{\gamma\ra 0}{ E}^i_{reg}\big(\!-\pa_i\vp_\gamma\big).
\end{align}
From \eref{dive} it follows indeed, see the equality in \eref{limj0},
\[
{ E}^i_{reg}\big(\!-\pa_i\vp_\gamma\big)=\frac12\,\mathcal{J}^0(\vp_\gamma)=\frac e2\,\chi_\gamma(\vec y(t_0)).
\]
As $\gamma\ra 0$ the r.h.s. of \eref{flu} gives  thus $e/2$ or $0$, according to whether at time $t_0$ the particle is in $V$ or not, which is the expected result\footnote{If $S$ does not intersect the singularity-string one can perform the limit in \eref{flu} trivially, obtaining the standard flux expression
\[
\Phi_S(t_0)=\int_S {E}^i_{reg}(t_0,\vec x)\,d\Sigma^i,
\]
which is then actually zero, since in this case the particle stays outside $V$.}.

We emphasize that the probably most striking feature of the result \eref{FB} is the vanishing of the Coulomb field -- that is, the field that for a {\it massive} particle in uniform motion carries the {\it entire} electric flux. We will come back to these issues, and to the physical meaning of \eref{FB}, in the concluding Section \ref{conclusions}.

\section{Li\'enard-Wiechert regularization\label{LWreg}}

Unfortunately for {\it unbounded} trajectories the ``regularized'' potential \eref{ABepsNoret} is not  a distribution. To see it notice that for large $r$ and large $t$ the integrands of \eref{ABepsNoret} and \eref{ABnoRet} become asymptotically identical and, consequently, as the latter functional for unbounded trajectories does  not represent a distribution (see the end of Section \ref{ABProof}), so does the former: both are indeed divergent.

To face the solution of Maxwell's equations for unbounded trajectories
we resort therefore to the  Li\'enard-Wiechert regularization,
introduced in Section \ref{limAUR}. Correspondingly we replace the light-like word-line $y^\m(\la)$ with the time-like world-line $Y^\m(\la)$ \eref{YVUR}, which -- when parametrized with time -- reads
\be
 Y^{\mu}(t)=(t,\vec{y}(Vt)), \quad\quad V<1.\label{YVB}
\ee
The corresponding Li\'enard-Wiechert field $F^{\mu\nu}$  $\eqref{FVdec}$ satisfies then the Maxwell equations \eref{MaxEqV} or, equivalently, the potential $A^\m$ \eref{AV} satisfies the equation \eref{boxAUR}. Since the current
$
J^{\mu}= e\int\delta^{4}(x-Y(\lambda))\,dY^\m
$
associated to the world-line \eref{YVB} satisfies the (trivial) distributional limit
\[
\LimV J^{\mu}= e\!\int\!\delta^{4}(x-y(\lambda))\,dy^\m =\mathcal{J}^{\mu},
\]
we conclude that, ${\it if }$ the field $\eqref{FVdec}$ admits a limit
\begin{equation}
\mathcal{F}^{\mu\nu}\equiv\LimV F^{\mu\nu}=
\LimV(C^{\m\n}+R^{\m\n})\equiv {\mathcal C}^{\m\n}+{\mathcal R}^{\m\n},
\label{FBFromLimV}
\end{equation}
the limiting  field $\mathcal{F}^{\mu\nu}$ is automatically a solution of the Maxwell equations  \eref{MaxEq}.
In writing \eref{FBFromLimV} we used that, if the field $F^{\mu\nu}$ admits a limit as a whole, the Coulomb and radiation fields will always admit limits separately\footnote{The reason for this is that, thanks to equations \eref{divCB}, where  ${\mathcal J}^\m $ and ${\mathcal K}^\m$   are well-defined distributions, possible ``divergent'' contributions to ${\mathcal C}^{\m\n}$ and ${\mathcal R}^{\m\n}$, {\it i.e.} distributions multiplying coefficients that diverge as $V\ra 1$, must satisfy {\it homogeneous} Maxwell equations. But since, on the other hand, these contributions must be supported on the world-line $y^\m(\la)$ and/or on the singularity surface $\Gamma^\m(\la,b)$, they are necessarily  zero.}, that we called ${\mathcal C}^{\m\n}$ and ${\mathcal R}^{\m\n}$.

\subsection{Distributional limits of fields and currents\label{limCB}}

If the limit \eref{FBFromLimV} exists,
important information about the fields ${\mathcal C}^{\m\n}$ and ${\mathcal R}^{\m\n}$ can be gained enforcing the distributional limits of equations \eref{CRVBianchi} and $\eqref{CRVMaxwell}$. A crucial step for this purpose is the determination of
the distributional limit of the ``current'' $K^\m$ defined in \eref{K}
\begin{equation}
\mathcal{K}^{\mu}\equiv\LimV K^{\mu}.\label{limK}
\end{equation}
In Appendix IV, see \eref{KB2}, we show that this limit exists  --  for bounded as well as unbounded trajectories -- its general expression being
\be
\mathcal{K}^{\mu}(\varphi)=e\!\int \! v^{\mu}(t-b)\,\varphi\big(t,b\vec{v}(t-b)+\vec{y}(t-b)\big)\,dt
\Big|_{b=0}^{b=\infty}.\label{KB1}
\ee
However, as we will see in the following,
the general form \eref{KB1} will give rise to different analytical  expressions according to  whether the trajectory is {\it bounded} or {\it unbounded}.

Applying the limit $V\ra 1$ to  equations \eref{CRVBianchi} and $\eqref{CRVMaxwell}$ we conclude  that the fields ${\mathcal C}^{\m\n}$ and  ${\mathcal R}^{\m\n}$ satisfy the Maxwell equations
\begin{equation}
\partial_{\mu}\mathcal{C}^{\mu\nu}=
\mathcal{J}^{\nu}+\mathcal{K}^{\nu},\quad\quad \partial_{[\mu}{\mathcal C}_{\nu\rho]}=0,\quad\quad
\partial_{\mu}\mathcal{R}^{\mu\nu}=-\mathcal{K}^{\nu},\quad\quad  \partial_{[\mu}{\mathcal R}_{\nu\rho]}=0.
\label{divCB}
\end{equation}
As observed previously, while the Li\'enard-Wiechert regularization has a clear physical meaning --  {\it i.e.} the ``regularized particle'' runs on the same orbit as the massless particle, but at a speed $V$ less than one -- it has the drawback of not being {\it manifestly} covariant;  nevertheless it will lead to manifestly Lorentz-covariant fields.

We stress, once more, that the entire procedure relies on a -- yet to be furnished -- proof that the limit \eref{FBFromLimV} exists.

\subsection{Coulomb and radiation fields for bounded motions}

For a bounded motion both our regularizations are available, so that we can use it to test and explore the power of the Li\'enard-Wiechert regularization.
Given \eref{ABnoRet1} and \eref{ABnoRet}, for a bounded motion we have trivially
\[
\LimV {A}^\mu= \mathcal A^\m,
\]
so that the existence of the limit \eref{FBFromLimV} -- see
\eref{FbFromA} --  is guaranteed.
We stress that this fact ensures that the fields $\mathcal{F}^{\mu\nu}$ derived with the two regularizations necessarily {\it coincide} -- equalling in fact \eref{FB} -- because they derive from the same potential \eref{ABnoRet}.

For a bounded trajectory the current \eref{KB1} gains no
contribution from $b=\infty$ -- because for $b\ra \infty$ the spatial argument of the test function tends to infinity -- while the contribution from $b=0$
gives
$$
\mathcal{K}^{\mu}(\varphi)=-e\!\int_{-\infty}^{\infty}v^\m\varphi(t,\vec{y}(t))\,dt=-{\mathcal J}^\m(\vp).
$$
This means that for a bounded motion we have the, {\it a priori} rather unexpected, result
\be\label{kmn1}
\mathcal{K}^{\mu}=-\mathcal{J}^{\mu}.
\ee From \eref{divCB} we see, indeed, that in this case the limiting Coulomb field $\mathcal{C}^{\mu\nu}$ satisfies {\it homogeneous} Maxwell equations and, actually, it vanishes, as we have already established in the framework of the covariant regularization in Section \ref{covreg}. Correspondingly the ``source'' of  the radiation field $\mathcal{R}^{\mu\nu}$ in \eref{divCB} is the total current $\mathcal{J}^{\mu}$.

\subsection{Vanishing of the Coulomb field\label{vcf}}

As a primary test of the Li\'enard-Wiechert regularization
we cross-check explicitly the disappearance of the Coulomb field
\be\label{cmn}
{\mathcal C}^{\m\n}=  \LimV C^{\mu\nu}=0,
\ee
where  $C^{\m\n}$ is the field $\eqref{CV}$  produced  by the regularized trajectory  \eref{YVB}. This will be instructive also because for an unbounded trajectory the limiting Coulomb field  will be no longer zero, and hence this test will allow us to better understand the mechanism of its appearance for such a trajectory -- where the covariant regularization is no longer available.

Starting from the world-line \eref{YVB} we set
\begin{equation}
\mathcal{V}^{\mu}(t)\equiv\frac{dY^{\mu}(t)}{dt}=\big(1,V\vec{v}(Vt)\big).\label{vmuV}
\end{equation}
Performing the analogous manipulations that led from \eref{FBeps} through \eref{fxe} and \eref{feps} to $\eqref{EBProof1}$, from $\eqref{CV}$ we find now
\begin{equation}
C^{\mu\nu}(\varphi)= \frac{(1-V^2)\,e}{4\pi}
\int\frac{X^{\mu}\mathcal{V}^{\nu}-X^{\nu}
\mathcal{V}^{\mu}}{r\left(r-V\vec{v}\cdot\vec{x}\right)^{2}}\,
\varphi\big(t+r,\vec{x}+\vec y\,\big)\,dtd^{3}x,\label{CBint}
\end{equation}
where $\vec v$ and $\vec y$ are evaluated at time $Vt$
and $X^{\mu}\equiv\left(r,\vec{x}\right)$. Performing again the change of variables \eref{bqVar} this integral turns into
\be
{C}^{\mu\nu} (\varphi)=\frac{(1-V^2)\,e}{4\pi}
\int\frac{X^{\mu}\mathcal{V}^{\nu}-X^{\nu}
\mathcal{V}^{\mu}}{r\left(r-Vb\right)^{2}}\,
\varphi\big(t+r,b\vec{v}+q_{a}\vec{N}_{a}+\vec{y}\,\big)d^{2}qdtdb.\label{VRegProof3}
\ee
For $b<0$ the denominator $(r-Vb)^2$ never vanishes, so that restricted to this region the integral multiplying $(1-V^2)$ in \eref{VRegProof3} converges in the limit $V\ra1$. Restricted to this region we have therefore  $\lim_{V\ra1} C^{\mu\nu}(\vp)\big|_{b<0}=0$.

In the complementary region $b\ge0$, as $V\ra1$, for small values of $q_a$
the denominator vanishes as $(r-Vb)^2\ra (r-b)^2\approx (q^2)^2/4b^2$. Accordingly for $b\ge0$ we perform the rescaling $q_{a}\rightarrow\sqrt{1-V^{2}}\,q_{a}$
getting
\be\label{cmn0}
{C}^{\mu\nu}(\varphi)\big|_{b\ge0}=\frac{e}{4\pi}\!\int_{b\ge0}\frac{X^{\mu}
\mathcal{V}^{\nu}-X^{\nu}\mathcal{V}^{\mu}}{r(q^{2}+b^{2})^{2}}
\,(r+Vb)^{2}\,\varphi\big(t+r,b\vec v+\!\sqrt{1-V^{2}}\,q_{a}\vec{N}_{a}+\vec{y}\,\big)
d^{2}qdtdb,
\ee
where now
\begin{equation}
r=\sqrt{b^2+(1-V^{2})q^{2}},\quad\quad X^{\mu}=\big(r,b\vec{v}+\sqrt{1-V^{2}}\,q_{a}\vec{N}_{a}\big).\label{VRegProof2}
\end{equation}
At this point we can use again the dominated convergence theorem (Appendix I) to
take in \eref{cmn0} the limit $V\ra 1$ inside the integral.
Since from \eref{vmuV} and $\eqref{VRegProof2}$ for $b\ge0$ we get
$\lim_{V\rightarrow1}X^{\mu}=b\left(1,\vec{v}\right)=bv^{\mu}
$ and $\lim_{V\rightarrow1}{\mathcal V}^{\mu}=v^\m$, it follows that
\[
\lim_{V\rightarrow1}
(X^{\mu}\mathcal{V}^{\nu}-X^{\nu}\mathcal{V}^{\mu})=0.
\]
Consequently we have
\[
\lim_{V\ra1} C^{\mu\nu}(\vp)=0
\]
for all $\vp$, which amounts to \eref{cmn}.

Let us underline the point in the proof where we used that the  trajectory is {\it bounded}. As $V\ra 1$ the test function in \eref{cmn0}, that cures the large distance divergences of the rest of the integrand, goes over in
$\varphi(t+b,\vec y(t)+b\vec v(t))$. If the trajectory were unbounded in the {\it past}, for example asymptotically linear, {\it i.e.} $\vec y(t)\approx \vec v_\infty t$ for $t\ra -\infty$, then the test function for large negative $t$ would take the {\it translation invariant} form $\varphi(t+b,(t+b)\vec v_\infty)$, implying that the large distance divergences of the integrand could no longer be cured; consequently   no function dominating uniformly the integrand -- needed for the validity of the dominated convergence theorem -- could be found.

Considering, on the other hand, the distributional limit for $V\ra 1$ of the radiation field \eref{RV}, by construction we get back the field \eref{FB}
\be\label{rmn}
\mathcal{R}^{\mu\nu}= \LimV R^{\m\n}=\frac{1}{2}\,Q^{\mu\nu}+\mathcal{P}(\mathcal{F}^{\mu\nu}_{reg}).
\ee
Due to the relevance that this limit will acquire in the case of unbounded trajectories, we perform it explicitly in Appendix VI.

\section{Unbounded accelerated motion\label{ubm}}

In this section we consider a massless  particle performing an unbounded motion, that in the infinite past is {\it free}. This is a common, and realistic, motion since all physically realizable macroscopic electromagnetic fields vanish at infinity, being actually of compact support. Correspondingly we suppose that the trajectory approaches for large negative times sufficiently fast a straight line ${\cal L}$ -- say $\vec y_\infty(t)=\vec v_\infty t$ -- the constant asymptotic velocity being obviously constrained by $|\vec v_\infty|=1$.  More precisely, setting
\be\label{asym}
\vec y(t)=\vec v_\infty t+ \vec \Delta(t),
\ee
we impose that -- shifting in case the origin of time -- for $t<0$ we have
\be
\big|\vec\Delta(t)\big|< \frac a{|t|^2},\quad\quad \big|\dot{\vec \Delta}(t)\big|< \frac b{|t|^2},\quad\quad
\big|\ddot{\vec \Delta}(t)\big|< \frac c{|t|^3},\label{limUBv}
\ee
where $a$, $b$ and $c$ are positive constants. This means that for $t\ra-\infty$ the kinematical quantities $\vec y(t)$, $\vec v(t)$ and $\vec a(t)$ fall off to their asymptotic values -- respectively  $\vec v_\infty t$, $\vec v_\infty$ and $0$ -- according to an inverse power law. In the following
we denote the constant asymptotic four-velocity (w.r.t. time) with
\begin{equation}
v_{\infty}^{\mu} \equiv(1,\vec v_{\infty}).
\end{equation}
We do not require any particular behavior of the trajectory for $t\ra+\infty$, as it will have no qualitative influence on the form of the electromagnetic field.

The main qualitative new feature of such an unbounded trajectory, with respect to a bounded one, is the modified form of the  singularity-string $\vec\gamma(t,b)$ \eref{gamma}: its end points at fixed time $t$ are now
\[
\vec\gamma(t,0)=\vec y(t),\quad\quad \vec\gamma(t,\infty)=\vec{v}_{\infty}t,
\]
and hence, contrary to the bounded case, the string has a {\it finite} extension. Consequently
during time evolution the singularity-string sweeps
out a surface whose  boundary is composed by the world-line of the particle
and by the world-line of a  {\it virtual} massless  particle traveling along the straight $\mathcal{L}$
\be\label{ymui}
y^\m_\infty(\la)=\la  v_\infty^\m.
\ee
Correspondingly, according to {\it Poincar\'e-duality}, the divergence of the field $Q^{\m\n}$ in \eref{pmn}, supported on the singularity surface, satisfies -- instead of \eref{dpmn} -- the modified equation
\be\label{upmn}
\pa_\m Q^{\m\n}={\mathcal J}^\n-{\mathcal J}_{\mathcal L}^\n,
\ee
 where ${\mathcal J}_{\mathcal L}^\m$ is the current associated to the world-line \eref{ymui}
\be\label{jl}
{\mathcal J}_{\mathcal L}^\m(x)=e\!\int\! v^\m_\infty\, \dl^4(x-\la v_\infty)\,d\la.
\ee

\subsection{The potential\label{cop}}

Since for an unbounded trajectory the trial functional \eref{ABnoRet} is not a distribution, to construct a well-defined potential we rely now on the Li\'enard-Wiechert regularization.

According to \eref{YVB} we introduce the regularized time-like world-line $Y^\m(t)=(t,\vec y(Vt))$ and consider the related  Li\'enard-Wiechert potential \eref{ABnoRet1}, {\it i.e.}
\begin{alignat}{1}
A^{\mu}(\varphi) & =\frac{e}{4\pi}\int\frac{\mathcal{V}^{\mu}(t)}{r}\,\varphi\big(t+r,\vec{x}+\vec{y}(Vt)\big)\,d^{4}x,
\quad\quad \mathcal{V}^\m(t)=(1,V\vec v(Vt)).
\label{AUBV}
\end{alignat}
This potential, however, does {\it not} converge  as $V\ra 1$ in the distributional sense, in that the candidate limit \eref{ABnoRet}  for unbounded trajectories is not a distribution. To overcome this difficulty
we introduce a ``compensating''  Li\'enard-Wiechert potential  $A_\infty^{\mu}$, produced by a {\it virtual} particle traveling along $\mathcal{L}$ with speed $V$, {\it i.e.} with world-line $Y^\m_\infty(t)=(t,V\vec v_\infty t)$, given by (see $\eqref{AURV}$)
\begin{equation}\label{aasym}
A^{\mu}_\infty(x)=\frac{e}{4\pi}\frac{\cal V^{\mu}_\infty}{\sqrt{({\cal V}_\infty x)^{2}-x^{2}{\cal V}^{2}_\infty}}, \quad\quad \mathcal{V}_{\infty}^{\mu}\equiv(1,V\vec{v}_{\infty}).
\end{equation}
As a distribution it is given by
\be
A_\infty^{\mu}(\varphi)  =\frac{e}{4\pi}\int\frac{\mathcal{V}_{\infty}^{\mu}}{r}\,
\varphi\big(t+r,\vec{x}+V\vec{v}_{\infty}t\big)\,d^{4}x.\label{AUBVr}
\ee
The world-lines $Y^\m(t)$ and $Y^\m_\infty(t)$ entail the same ``pathological'' behavior as $t\ra-\infty$, but from Section \ref{limAUR} we know how to cure the pathologies of the latter: before taking the limit $V\ra1$ we must perform a gauge transformation with gauge function \eref{gaugeT}, that in the present case becomes
\[
\Lambda(x)=\frac{e}{4\pi}\ln\left|(\mathcal{V}_{\infty}x)
-\sqrt{(\mathcal{V}_{\infty}x)^{2}-  x^{2}\mathcal{V}_{\infty}^2 }\,\right|.
\]
This suggests to a define a potential as
\be\label{unb}
{\mathcal A}^\m\equiv \LimV(A^{\mu}+\pa^\m\Lambda)
\ee
and, indeed, we can prove the following theorem.
\vskip0.2truecm\noindent
{\bf Theorem.} {\it If the unbounded trajectory $\vec y(t)$ for $t\ra-\infty$ entails the asymptotic flatness conditions \eref{limUBv}, the distributional limit \eref{unb} exists and is given by}
\begin{align}\label{unbp1}
{\mathcal A}^\m(\vp)=&\frac{e}{4\pi}\int\!\frac1r\,\Big(v^{\mu}(t)\,
\varphi\big(t+r,\vec{x}+\vec{y}(t)\big)-v_{\infty}^{\mu}\,
\varphi\big(t+r,\vec{x}+\vec{v}_{\infty}t\big)\!\Big)d^{4}x\\
&+\frac e{2\pi} \int\! \frac {x^\m}{x^2}\,H(v_\infty x)\,\vp(x)\,d^4x,\label{unbp2}
\end{align}
where the principal-part integration prescription for the factor $1/x^2=1/(x^\n x_\n)$ is understood.
\vskip0.2truecm
\noindent {\it Proof}. We  write
\[
A^{\mu}+\pa^\m\Lambda=A_{1}^{\mu}+A_{2}^{\mu},\quad\quad\mbox{where}\quad
A_{1}^{\mu}\equiv A^{\mu}-A_\infty^{\mu},\quad\quad A_{2}^{\mu}\equiv A_\infty^{\mu}+\partial^{\mu}\Lambda.
\]
As we have shown in Section \ref{limAUR}, the limit of $A_{2}^{\mu}$ as $V\ra 1$ in
the sense of distributions  exists and is given by $\eqref{AUR}$
-- the potential generated by a  particle in light-like linear motion along $\mathcal{L}$, with world-line \eref{ymui}. This limit corresponds hence to the term \eref{unbp2}.
 As for  $A_{1}^{\mu}$
from $\eqref{AUBV}$ and $\eqref{AUBVr}$ we get
\begin{equation}
A_{1}^{\mu}(\varphi)=\frac{e}{4\pi}\int\!\frac1r \,\Big(\mathcal{V}^{\mu}(t)\,
\varphi\big(t+r,\vec{x}+\vec{y}(Vt)\big)-\mathcal{V}_{\infty}^{\mu}\,
\varphi\big(t+r,\vec{x}+V\vec{v}_{\infty}t\big)\!\Big)d^{4}x.\label{A1}
\end{equation}
As we will prove in Appendix V, when considering the limit of this expression for $V\ra 1$ we can swap the limit with the integral sign and the result is therefore simply \eref{unbp1}. The main point of the proof is that, as the two terms in \eref{A1} have the {\it same} singular behavior
as $r\rightarrow\infty$ (along the direction of $\vec v_\infty$) and simultaneously $t=-r\rightarrow-\infty$,
in the limit $V\rightarrow1$ the divergences cancel out from their difference -- if the asymptotic conditions \eref{limUBv} are satisfied.   \hfill $\square$
\\

As in the case of a linear motion it is again possible to construct a potential $\mathcal{A}'^\mu$ satisfying the Lorenz gauge.  The four-divergence of the term \eref{unbp1} is, in fact, zero, since it is the distributional limit for $V\rightarrow 1$ of the potential $A_{1}^{\mu}$ that -- being the difference of two potentials obeying the Lorenz gauge --  satisfies $\partial_{\mu}A_{1}^{\mu}=0$.  The term  \eref{unbp2} corresponds to the potential \eref{AUR} of a massless particle in linear motion with four-velocity $v_{\infty}$ and has, instead, a non-vanishing divergence. In order to restore the Lorenz gauge it is thus sufficient to perform the same gauge transformation \eref{lorg} of Section \ref{limAUR} , this time in the form
\begin{equation}
\mathcal{A}'^\mu=\mathcal{A}^\mu-\partial^{\mu}\!\left(\frac{e}{4\pi}H(v_{\infty}x)
\ln|x^{2}|\right),\quad\quad  \pa_\m\mathcal{A}'^\mu=0. \label{LGaugeUB}
\end{equation}

\subsection{Solution of Maxwell's equations}

According to \eref{unb} the field strength  $\mathcal{F}^{\mu\nu}=
\partial^{\mu}\mathcal{A}^{\nu}-\partial^{\nu}\mathcal{A}^{\mu}$ can be computed
as the limit under $V\ra 1$ of the field $F^{\mu\nu}=\partial^{\mu}A^{\nu}-\partial^{\nu}A^{\mu}$,
with $A^{\mu}$  the Li\'enard-Wiechert potential $\eqref{AUBV}$. This means that we have, once more,
\begin{equation}\label{limun}
\mathcal{F}^{\mu\nu}=\LimV F^{\mu\nu}= \LimV(C^{\m\n}+R^{\m\n})={\mathcal C}^{\m\n}+{\mathcal R}^{\m\n},
\end{equation}
where $F^{\mu\nu}$ is again the Li\'enard-Wiechert field $\eqref{FVdec}$-\eref{RV} relative to the regularized trajectory \eref{YVB}. What we have shown in the previous section is, actually, that the limit \eref{limun} exists -- although the (apparently natural) potential \eref{AUBV} does not admit a distributional limit. To evaluate the limit \eref{limun} explicitly  we  consider the
Coulomb and radiation fields  separately. Useful information will then be gained again  from the Maxwell equations \eref{divCB}, that we know to be satisfied automatically by the limiting fields ${\mathcal C}^{\m\n}$ and ${\mathcal R}^{\m\n}$.

\subsubsection{Currents\label{loc}}

As shown in Appendix IV,
the ``current'' $\mathcal{K}^\m$ showing up in \eref{divCB} has the general expression
\eref{KB1}. But now -- contrary to the bounded case -- by virtue of \eref{asym} and the asymptotic behaviors $\eqref{limUBv}$ we have
\begin{equation}
\lim_{b\rightarrow\infty}\big(\vec{y}(t-b)+b\vec{v}(t-b)\big)=
\lim_{b\rightarrow\infty}\big((t-b)\vec{v}_{\infty}+b\vec{v}_{\infty}\big)=
t\vec{v}_{\infty}.\label{limgammaUB}
\end{equation}
Consequently both  terms in \eref{KB1} survive and we obtain (see \eref{jl})
 \be
\mathcal{K}^{\mu}(\varphi)  =e\!\int\! v_{\infty}^{\mu}\,\varphi(t,t\vec{v}_{\infty})\,dt-e\!\int\! v^{\mu}(t)\,\varphi(t,\vec{y}(t))\,dt=
 \mathcal{J}_{\mathcal{L}}^{\mu}(\varphi)-\mathcal{J}^{\mu}(\varphi).
\ee
We have thus
\be\label{dRUB}
\mathcal{K}^{\mu}=  \mathcal{J}^{\mu}_{\mathcal L}- \mathcal{J}^{\mu}.
\ee
Equations \eref{divCB} imply then
that the sources of the Coulomb and radiation fields are both non vanishing, being given by
\be\label{maxcr}
\partial_{\mu}\mathcal{C}^{\mu\nu}=\mathcal{J}_{\mathcal{L}}^{\nu}, \quad\quad
\partial_{\mu}\mathcal{R}^{\mu\nu}=  \mathcal{J}^{\nu}-   \mathcal{J}_{\mathcal{L}}^{\nu}.
\ee
For the particular case of a particle in linear motion along $\mathcal{L}$ we have $\mathcal{J}^\m= \mathcal{J}_{\mathcal{L}}^\m$, and equations \eref{maxcr} are thus in agreement with the shock-wave solution \eref{FUR}, for which we have indeed $\mathcal{R}^{\mu\nu}=0$.

\subsubsection{Electromagnetic field}

To determine the solution of Maxwell's equations explicitly it remains to evaluate the limits \eref{limun} for an unbounded motion where, we recall, the fields $C^{\m\n}$ and $R^{\m\n}$ are given in \eref{CV}, \eref{RV}.

The limit of the radiation field $R^{\m\n}$ can be evaluated in exactly the same way as  for the bounded motion, see Appendix VI. The only delicate point is again the swapping of the limit $V\ra 1$ with the integrals, but this time this can be done thanks to the bound  \eref{limUBv} on the acceleration $\vec a(t)=\ddot{\vec \Delta}(t)$. The result for $\mathcal{R}^{\mu\nu}$ is thus again \eref{rmn}.

For what concerns the limit of the Coulomb field $\mathcal{C}^{\mu\nu}$ we observe that, from \eref{divCB} and \eref{maxcr},  it must satisfy the equations
\[
\partial_{\mu}\mathcal{C}^{\mu\nu}=\mathcal{J}_{\mathcal{L}}^{\nu},\quad\quad
\partial_{[\mu}{\mathcal C}_{\nu\rho]}=0,
\]
where the current $\mathcal{J}_{\mathcal{L}}^\m$ is that of a {\it virtual} particle moving with constant light-like four-velocity $v^\m_\infty$ along $\mathcal{L}$. For uniqueness reasons
$\mathcal{C}^{\mu\nu}$ is then the  shock-wave field \eref{FUR}, that is
\be\label{csw}
 \mathcal{C}^{\mu\nu}=\frac{e}{2\pi}\frac{v^{\mu}_\infty x^{\nu}-v^{\nu}_\infty x^{\mu}}{x^{2}}\,\delta
(v_\infty x).
\ee
In conclusion, a light-like particle following an unbounded motion that is asymptotically linear in the past, creates the electromagnetic field
\begin{equation}
\mathcal{F}^{\mu\nu}=\mathcal{P}(\mathcal{F}^{\mu\nu}_{reg})+  \frac{1}{2}\,Q^{\mu\nu}
+\mathcal{C}^{\mu\nu}.
\label{Fug}
\end{equation}

The first term is a regular distribution, that corresponds essentially to the point-wise limit of the Li\'enard-Wiechert radiation field; the second term is a $\dl$-function supported on the dynamical singularity-string \eref{gamma}, that has now, however, a finite extension. The field $\mathcal{P}(\mathcal{F}^{\mu\nu}_{reg})$ diverges on the same singularity-string, too.
The  term $\mathcal{C}^{\mu\nu}$ --  a remnant of the Coulomb field -- is a $\dl$-function supported on a plane shock-wave that moves at the speed of light. We stress that all these fields -- being the limits of (causal) Li\'enard-Wiechert fields -- respect automatically {\it causality}, representing thus phenomena that propagate at the speed of light. In particular all points of the singularity-string \eref{gamma} can be seen to move at the speed of light, see \cite{AL}.

From equations \eref{upmn} and \eref{maxcr} it follows that the currents carried by the three fields in \eref{Fug} are (compare with \eref{demo1} and \eref{demo2} of the bounded case)
\begin{align}
\label{curd1}
\pa_\m\!\left(\!\frac{1}{2}\,Q^{\mu\nu}\!\right)&=
 \frac12\,({\mathcal J}^\n-{\mathcal J}_{\mathcal L}^\n),\\
\pa_\m\big(\mathcal{P}(\mathcal{F}^{\mu\nu}_{reg})\big)&=
 \frac12\,({\mathcal J}^\n-{\mathcal J}_{\mathcal L}^\n),\label{curd2}\\
\pa_\m \mathcal{C}^{\mu\nu}&={\mathcal J}_{\mathcal L}^\n,
\end{align}
and obviously sum up to give $\mathcal{J}^\n$. Only the sum of the first two fields satisfies the Bianchi identity, while $\mathcal{C}^{\mu\nu}$ satisfies it independently.

With respect to a bounded motion the new ingredient in \eref{Fug} is the (re)appearance of the shock-wave -- a feature that might by itself seem unexpected. The reason for the presence of this shock-wave is clearly that the particle in the infinite past approaches a straight line: although not being strictly linear, the almost infinite duration of the linear motion compensates the never strictly vanishing acceleration, giving rise to a kind of coherent ``non-perturbative'' shock-wave -- {\it as if} the particle moved in the infinite past on a straight line. Obviously the presence of this field is also needed to give rise to a consistent solution of Maxwell's equations. In fact, since the boundary of the singularity surface is given by the world-line $y^\m(\la)$ {\it and} by ${\mathcal L}$, the divergence of the first two terms in \eref{Fug} equals ${\mathcal J}^\n-{\mathcal J}_{\mathcal L}^\n$, rather than ${\mathcal J}^\n$: the shock-wave is precisely needed to cure this mismatch. With this respect equation \eref{curd2} may look rather unexpected -- remember that $\mathcal{P}(\mathcal{F}^{\mu\nu}_{reg})$ corresponds to the {\it naiv} limit of the radiation field, often presented in the literature  erroneously  as the ``true'' field of a massless particle --  and we give an independent proof of it in Appendix VIII.

A peculiar feature of the field \eref{Fug} is that it
vanishes ``in front of'' the shock-wave \eref{csw}, {\it i.e.} for fixed $t$ it vanishes for any $\vec x$ such that $\vec x\cdot\vec v_\infty>t$.
For the field $Q^{\m\n}/2$, supported on the singularity surface,  this is obvious since for fixed $t$ the singularity-string \eref{gamma} starts from the particle's position and ends on the center $\vec v_\infty t$ of the shock-wave, lying always behind it.
For the regular field $\mathcal{F}^{\mu\nu}_{reg}$ \eref{RBpointWise}
this follows instead from the fact that for points such that $\vec x\cdot\vec v_\infty>t$, the retarded-time conditions \eref{llcaus} admit no solution for $\la$ (for the proof see Appendix VII). The reason for this is that the total field begins its life in the infinite past, originating from a linear motion that produces first of all a shock-wave propagating linearly at the speed of light. Since the field $\mathcal{F}^{\mu\nu}_{reg}$ -- a true radiation field -- arises from acceleration, this field is thus created ``after'' the shock-wave and propagates non-linearly: it can therefore never reach the shock-wave and lies, hence, always behind it.

\subsubsection{Shock-wave in an idealized trajectory}

Since the appearance of the shock-wave \eref{csw} represents an important  new physical feature of the solution \eref{Fug} for an unbounded trajectory -- for a bounded trajectory in Section \ref{vcf} we showed explicitly how in the
Li\'enard-Wiechert regularization the Coulomb field vanishes  -- to achieve a more direct understanding of the reappearance of this field, we derive it explicitly as the limit for $V\ra 1$ of \eref{VRegProof3} for an {\it idealized} trajectory, where the conditions \eref{limUBv} are ``strongly'' satisfied, {\it i.e.} for a trajectory that is a straight line until a certain time, say $t=0$,
\be\label{ylin}
\vec y(t)= \vec v_\infty t, \quad\quad\mbox{for}\quad t\le0.
\ee
We start again from the general expression \eref{VRegProof3}
\be
{C}^{\mu\nu} (\varphi)=\frac{(1-V^2)\,e}{4\pi}
\int\frac{X^{\mu}\mathcal{V}^{\nu}-X^{\nu}
\mathcal{V}^{\mu}}{r\left(r-Vb\right)^{2}}\,
\varphi\big(t+r,b\vec{v}+q_{a}\vec{N}_{a}+\vec{y}\,\big)d^{2}qdtdb,\label{VRegProof3a}
\ee
where, we recall, $X^\m=(r,b\vec v+q_a\vec N_a)$, $r=\sqrt{b^2+q^2}$, $\mathcal{V}^{\m}=(1,V \vec v)$ and the variables $\vec y$, $\vec v$ and $\vec N_a$ are evaluated at time $Vt$.

We analyze the limit under $V\ra 1$ of this integral considering separately the regions $t>0$ and $t<0$. For $t>0$ we may proceed in the same way as in Section \ref{vcf} -- even if the trajectory is unbounded in the future -- since, as we explained at the end of that section, the procedure fails only for trajectories unbounded in the past. For $t>0$ the limit of \eref{VRegProof3a} for $V\ra 1$ is thus again zero.

For $t<0$ the trajectory \eref{ylin} is linear and we proceed in a different way, resembling in some sense the procedure that allowed in Section \ref{secDistLimR} to derive the shock-wave for an {\it infinite} linear motion. We perform the change of variables
$(t,b)\ra (t',b')$
\[
t'=t+r,\quad\quad
b'=\frac{b-Vr}{\sqrt{1-V^{2}}},
\]
with inverse transformations
\[
b=\frac{b'+VR}{\sqrt{1-V^{2}}},\quad\quad t=t'-\frac{R+Vb'}{\sqrt{1-V^2}},\quad\quad\mbox{where}\,\,\, R\equiv\sqrt{b'^2+q^2}.
\]
In particular we have
\[
r=\frac{R+Vb'}{\sqrt{1-V^{2}}},  \quad\quad r-Vb=\sqrt{1-V^2}R,\quad\quad
\frac{\partial b}{\partial b'}=\frac{r}{R}.
\]
The variable $t'$ ranges now from $-\infty$ to $r$, while $b'$ ranges again from $-\infty$ to $\infty$.
Inserting these elements, as well as \eref{ylin}, in \eref{VRegProof3a}, we obtain  (for $t<0$ the vectors $\vec N_a\perp \vec v_\infty$ are time-independent)
\begin{equation}
{C}^{\mu\nu}(\varphi)\big|_{t<0}=\frac{e}{4\pi}\int_{t'<r}
\frac{\left(X'^{\mu}\mathcal{V}^{\nu}-X'^{\nu}\mathcal{V}^{\mu}\right)}
{R^{3}}\,\varphi\big(t',q_{a}\vec{N}_{a}+\big(Vt'-b'\sqrt{1-V^{2}}\,\big)\vec{v}_{\infty}\big)
d^{2}qdt'db',\label{idealProof4}
\end{equation}
where
\[
X'^{\mu}= \Big(\sqrt{\frac{1-V}{1+V}}\,(R-b'),q_{a}\vec{N}_{a}\Big).
\]
Since for $V\ra 1$ we have $r\ra \infty$, swapping in \eref{idealProof4} the limit with the integral, and omitting the primes, we get
\[
\mathcal{C}^{\mu\nu}(\varphi)=\lim_{V\ra1} C^{\mu\nu}(\vp)
=\frac{e}{4\pi}\int\frac{m{}^{\mu}v_\infty^{\nu}-m{}^{\nu}
v_\infty^{\mu}}{(b^2+q^2)^{3/2}}\,\varphi\big(t,t\vec{v}_{\infty}+q_{a}\vec{N}_{a}\big)d^{2}qdtdb,
\]
where $m^{\mu}\equiv(0,q_{a}\vec{N}_{a})$. Integrating
over $b$ we obtain eventually
\[
\mathcal{C}^{\mu\nu}(\varphi)=\frac{e}{2\pi}\int\frac {{m{}^{\mu}v_\infty^{\nu}-m{}^{\nu}
v_\infty^{\mu}}}{q^2}\,\varphi\big(t,t\vec{v}_{\infty}+q_{a}\vec{N}_{a}\big)d^{2}qdt,
\]
in agreement with \eref{csw}.

Notice that for a trajectory satisfying \eref{ylin} the terms
$\mathcal{P}(\mathcal{F}^{\mu\nu}_{reg})+ Q^{\mu\nu}/2$ of \eref{Fug} for $t<0$ vanish, so that for negative times the field is a pure shock-wave.

\section{Conclusions and interpretation} \label{conclusions}

We have shown that Maxwell's equations for massless particles -- traveling at the speed of light -- admit exact and explicit solutions in the space of distributions.  The corresponding fields are uniquely determined and respect causality: they represent, therefore, the correct generalizations of the Li\'enard-Wiechert fields from time-like to light-like trajectories.

Unlike the time-like case, bounded and unbounded light-like motions produce radically different electromagnetic fields. With respect to the field \eref{FB} of a bounded motion, the field \eref{Fug} of an unbounded motion is, in a certain sense, less singular. The former contains  in fact a $\dl$-function $Q^{\m\n}$ supported at fixed  time on an infinitely extended singularity-string, while the singularity-string of the latter has a finite extension. This feature can be interpreted observing that the former is produced by a particle that is bounded for all times and hence eternally accelerated: the field keeps memory of this acceleration and reaches at each time in each direction spatial infinity. Notice, in particular, that the regular field $\mathcal{P}(\mathcal{F}^{\mu\nu}_{reg})$ in \eref{FB} -- that encodes, properly speaking, the radiation -- entails the same singularity locus as  $Q^{\m\n}$.

On the contrary a particle performing an unbounded motion is  {\it free} in the infinite past, where it does not produce radiation, and correspondingly the singularity locus
of  $\mathcal{P}(\mathcal{F}^{\mu\nu}_{reg})+ Q^{\mu\nu}/2$
is at each time a bounded region, {\it i.e.} a bounded string. In addition to this field \eref{Fug} entails a shock-wave -- at fixed time a $\dl$-function on a two-dimensional surface -- and in this sense the total field is more singular than the field of a bounded motion. In some sense the shock-wave represents the counterpart of the Coulomb field of a massive particle, in that it does not carry radiation.

We constructed for all physical situations considered -- linear, bounded and unbounded accelerated motions -- well-defined four-potentials, a more delicate task since potentials are in general more singular then the field strengths. While for bounded motions the potential can be constructed as the point-wise limit of the potential of a time-like trajectory, for unbounded motions before taking the (distributional) limit one must consider an appropriate gauge transformation. In this way for unbounded motions one obtains a potential that does no longer satisfy the Lorenz-gauge $\pa_\m \mathcal{A}^\m =0$, which can, however, be restored by the gauge transformation \eref{LGaugeUB}

While the mathematical significance of our results is transparent, their physical interpretation and consequences are largely to be investigated. In particular there are some peculiar aspects of the structure of our basic results \eref{FB} and \eref{Fug}, that call for a physical explanation. One such feature is the ``democratic'' partition of the total electric flux between the two fields of \eref{FB}, see equations \eref{demo1}, \eref{demo2}: the first field is {\it semi-integer}, in the sense that its flux through an {\it arbitrary} surface, closed or not, equals  $Ne/2$ with $N\in \mathbb{Z}$,  while the flux of the second field through an arbitrary {\it closed} surface equals $0$ or $e/2$. It would be rather surprising if these features had no direct physical counterpart, still to be discovered.

Once the electromagnetic field of a massless particle in generic motion is known, one can eventually face the {\it radiation problem} (for a preliminary analysis see \cite{Y}). To settle this issue one must first of all construct a {\it distribution-valued energy-momentum tensor}\footnote{The techniques to construct such a tensor developed in \cite{LPAM} for massive particles should apply also to massless ones.} of the electromagnetic field, $\mathcal{T}^{\m\n}_{em}$, that in the complement of the singularity locus $\Sigma$ -- {\it i.e.} the singularity surface \eref{Gamma} and the shock-wave position $\vec v_\infty \cdot \vec x=t$ -- equals the standard energy-momentum tensor:
\[
\mathcal{T}^{\m\n}_{em}\big|_{\mathbb{R}^4 \backslash \Sigma}=\mathcal{F}^{\m\a}
\mathcal{F}_\a{}^\n+\frac14\,\eta^{\m\n}\mathcal{F}^{\a\beta}\mathcal{F}_{\a\beta}.
\]
Once such a tensor has been constructed one can determine its distributional divergence $\pa_\m\mathcal{T}^{\m\n}_{em}$, and check whether there exists an equation of motion for a time-like world-line $y^\m(\la)$, such that the total energy-momentum tensor
$\mathcal{T}^{\m\n}=\mathcal{T}^{\m\n}_{em}+\int\! u^\m u^\n\dl^4(x-y(\la))\,d\la$ is conserved: $\pa_\m\mathcal{T}^{\m\n}=0$. If no such equation of motion exists, we must conclude that {\it classical Electrodynamics} of massless charged particles is inconsistent, in that energy is not conserved. This would imply, in particular, that the analysis of the ``emitted radiation'' of such a particle is meaningless, since the emitted energy would not equal the energy lost by the particle. In this way
the results of the present paper open the possibility to check whether classical Electrodynamics of massless charged particles is consistent or not. This problem will be addressed elsewhere.

\paragraph{Acknowledgments.}

This work is supported in part by the INFN Iniziativa Specifica TV12
and by the Padova University Project CPDA119349.

\vskip0.5truecm

\section{Appendices}\label{app}

\subsection*{Appendix I}

{\bf Distributions.} The Schwarz space
of test functions  $\mathcal{S}$ is the set of  complex
functions $\varphi\in\mathcal{C}^{\infty}\left(\mathbb{R}^{4}\right)$
such that
\begin{equation}
\left\Vert \varphi\right\Vert _{\mathcal{P},\mathcal{Q}}<\infty,\quad\mbox{for all polynomials $\mathcal{P}$ and $\mathcal{Q}$},\label{testFuncProp}
\end{equation}
where
\[
\left\Vert \varphi\right\Vert _{\mathcal{P},\mathcal{Q}}\equiv \sup_{x\in\mathbb{R}^{4}}
\left|\mathcal{P}(x)\mathcal{Q}(\partial)\varphi\left(x\right)\right|
\]
are {\it semi-norms} of $\varphi$, and $\mathcal{P}$ and $\mathcal{Q}$
are generic polynomials of coordinates and derivatives respectively.
The space of {\it tempered distributions} $\mathcal{S}'$ is the space of linear
continuous functionals on $\mathcal{S}$. It can be shown that
these functionals are characterized by the property
\begin{equation}
F\in\mathcal{S}'\quad\Leftrightarrow\quad
\left|F(\varphi)\right|\leq\sum_{\mathcal{P},\mathcal{Q}}
 C_{\mathcal{P},\mathcal{Q}}\left\Vert \varphi\right\Vert _{\mathcal{P},\mathcal{Q}},\,\,\forall\varphi\in\mathcal{S},\label{distDef}
\end{equation}
where the sum must contain a {\it finite} number of terms and the coefficients $C_{\mathcal{P},\mathcal{Q}}$ must be independent of $\vp$.

In the text we use several times the following theorem that allows, through the corollary stated below, to interchange limits with integrals.

\vskip0.2truecm\noindent
{\bf Dominated convergence theorem.} {\it Let $\{f_n\}$ be a sequence of functions belonging to $L^1(\mathbb{R}^D)$. If there exists a function $f(x)$ and a positive function $g\in L^1(\mathbb{R}^D)$ such that
\begin{itemize}
\item[a)] $\lim_{n\ra\infty}f_n(x)=f(x)$,
\item[b)] $|f_n(x)|\le g(x),\quad\forall n,\,\forall x$,
\end{itemize}
then
\[
{L^1}\!-\!\lim_{n\ra\infty}f_n=f.
\]
}
\\
{\bf Corollary.} {\it If the sequence $f_n$ satisfies the hypotheses $a)$ and $b)$ of the dominated convergence theorem then
\[
\lim_{n\ra\infty}\int\! f_n(x)\,d^Dx=\int\! f(x)\,d^Dx,
\]
{\it i.e.} the limit can be swapped with the integral sign.}

\subsection*{Appendix II}

In this appendix we derive the limit $\eqref{LimJeps}$ with $\mathcal{J}_{\varepsilon}^{\mu}$
given in $\eqref{Jreg}$. Thanks to manifest Lorentz-invariance it
is sufficient to establish this limit for the time component $\mu=0$,
meaning that we must prove the ordinary limits, see $\eqref{LLCurrent}$,
\begin{equation}
\lim_{\varepsilon\rightarrow0}\mathcal{J}_{\varepsilon}^{0}(\varphi)=\mathcal{J}^{0}(\varphi)
=e\!\int\!\varphi(t,\vec{y}(t))\, dt\label{limj0}
\end{equation}
for an arbitrary test function $\varphi$.

By construction the expression $\eqref{Jreg}$ is invariant under reparametrization of the world-line, so that we can choose as parameter $\lambda= y^{0}(\lambda)\equiv t$. In this way the kinematic
quantities simplify to $u^{\mu}=(1,\vec{v})$, $w^{\mu}=(0,\vec{a})$ and $b^\m=(0,\dot{\vec{a}}\,)$.
Using  the identity  \eref{fxe} and proceeding as in \eref{feps} and \eref{EBProof1},
the application of the time component of $\eqref{Jreg}$ to a test function
reads
\be\nn
\mathcal{J}_{\varepsilon}^{0}(\vp) =\frac{\varepsilon^{2}e}{4\pi}\int\frac{3(\vec{a}\cdot\vec{x})^{2}+
(r_{\varepsilon}-\vec{v}\cdot\vec{x})(\dot{\vec{a}}\cdot\vec{x})}
{r_{\varepsilon}\left(r_{\varepsilon}-\vec{v}\cdot\vec{x}\right)^{4}}
\,\varphi(t+r_{\varepsilon},\vec{x}+\vec{y})\,d^3xdt,
\ee
where the kinematic quantities $\vec y$, $\vec v$ and $\vec a$ are evaluated at time $t$ and $r_\ve=\sqrt{r^2+\ve^2}$. Performing the change of variables \eref{bqVar} $\vec{x}=b\vec{v}+q_{a}\vec{N}_{a}$ this expression turns into
 \be
\mathcal{J}_{\varepsilon}^{0}(\vp) =
 \frac{\varepsilon^{2}e}{4\pi}\int\frac{3\big(q_{c}\vec{N}_{c}\cdot\vec{a}\,\big)^{2}\!
+(r_{\varepsilon}-b)\big(q_{c}\vec{N}_{c}\cdot
\dot{\vec{a}}-ba^{2}\big)}{r_{\varepsilon}(r_{\varepsilon}-b)^{4}}\,
\varphi(t+r_{\varepsilon},b\vec{v}+q_{c}\vec{N}_{c}+\vec{y})\,dtdbd^{2}q,\label{j02}
\ee
where now $r_{\varepsilon}=\sqrt{b^{2}+q^{2}+\varepsilon^{2}}$.
As $\varepsilon\rightarrow0$, in the region $b<0$ the integrand
develops no singularities around $q_{a}=0$ and the limit vanishes.
We can therefore restrict the integration over $b$ to the half line
$b>0$.

To perform the limit we write
\[
\frac{1}{r_{\varepsilon}-b}=\frac{r_{\varepsilon}+b}{q^{2}+\varepsilon^{2}}
\]
and rescale the variable $q_{a}\rightarrow\varepsilon q_{a}$. Taking
into account the rescaling of the numerator and of the measure $d^{2}q$
in $\eqref{j02}$, as $\varepsilon\rightarrow0$ the integral develops
\textit{a priori} terms that diverge as $1/\varepsilon^{2}$ and terms
that diverge as $1/\varepsilon$. By inspection one sees that the
poles $1/\varepsilon^{2}$ cancel each other out, while the poles $1/\varepsilon$
cancel individually thanks to symmetric integration over $q_{a}$.
These cancelations are obviously due to the fact that by construction
the limit $\eqref{LimJeps}$ exists. To compute the finite terms of
$\eqref{j02}$ as $\varepsilon\rightarrow0$, after the
rescaling $q_{a}\rightarrow\varepsilon q_{a}$, due to the prefactor
$\varepsilon^{2}$ we must expand the integrand in powers of $\varepsilon$
keeping only terms of order $1/\varepsilon^{2}$. After the expansion of
 the numerator and the denominator, as well as of the test
function $\varphi$, the resulting integrals over $q_{a}$ are elementary
and, using the identities $\eqref{qId}$, one gets
\[
\lim_{\varepsilon\rightarrow0}\mathcal{J}_{\varepsilon}^{0}(\varphi)=\frac{e}
{2}\int_{-\infty}^{\infty}dt
\int_{0}^{\infty}\!db\left(b^{3}a^{i}a^{j}\partial_{i}\partial_{j}+b^{2}(a^{2}\partial_{0}
+\dot{a}^{i}\partial_{i}+
a^{2}v^{i}\partial_{i})+2ba^{2}\right)\varphi(t+b,b\vec{v}+\vec{y}).
\]
Making repeated use of the identities ($\varphi\equiv\varphi(t+b,b\vec{v}+\vec{y})$)
\[
\frac{\partial\varphi}{\partial b}=(\partial_{0}+v^{i}\partial_{i})\varphi,\quad\quad\frac{\partial\varphi}{\partial t}=\left(\partial_{0}+(v^{i}+ba^{i})\partial_{i}\right)\!\varphi,
\]
and integrating several times by parts, one sees that the three terms
proportional to $a^{2}$ cancel each other, while the remaining two
can be rewritten as
\begin{alignat*}{1}
\lim_{\varepsilon\rightarrow0}\mathcal{J}_{\varepsilon}^{0}(\varphi) & =\frac{e}{2}\int_{-\infty}^{\infty}dt\int_{0}^{\infty}db\,(b^{3}a^{i}a^{j}
\partial_{i}\partial_{j}+b^{2}\dot{a}^{i}\partial_{i})\,\varphi(t+b,b\vec{v}+\vec{y})\\
 & =\frac{e}{2}\int_{-\infty}^{\infty}dt\int_{0}^{\infty}db\, b^{2}\!\left(\frac{\partial}{\partial t}-\frac{\partial}{\partial b}\right)\!\left(a^{i}\partial_{i}\varphi(t+b,b\vec{v}+\vec{y})\right)\\
 & =e\int_{-\infty}^{\infty}dt\int_{0}^{\infty}db\, ba^{i}\partial_{i}\varphi(t+b,b\vec{v}+\vec{y})\\
 & =e\int_{-\infty}^{\infty}dt\int_{0}^{\infty}db\left(\frac{\partial}{\partial t}-\frac{\partial}{\partial b}\right)\!\varphi(t+b,b\vec{v}+\vec{y})=e\int_{-\infty}^{\infty}\varphi(t,\vec{y})\,dt,
\end{alignat*}
which is $\eqref{limj0}$.

\subsection*{Appendix III}

In this appendix we derive the limit \eref{EBProof5} with ${\cal F}^{i0}_\ve(\vp)$ given in \eref{EBProof1}. Performing in the latter the change of variables \eref{bqVar}, and  using that the constraint $|\vec v|^2=1$ implies $\vec v\perp \vec a$, we obtain
\[
{\cal F}^{i0}_\ve
(\varphi)=\frac{e}{4\pi}\int\!\frac{\big(\vec{a}\cdot q_{c}\vec{N}_{c}\big)\big((b-r_{\varepsilon})v^{i}+q_{b}N_{b}^{i}\big)
-r_{\varepsilon}(r_{\varepsilon}-b)a^{i}}{r_{\varepsilon}
(r_{\varepsilon}-b)^{2}}\,\varphi(t+r_{\varepsilon},b\vec{v}+q_{a}\vec{N}_{a}
+\vec{y})\,dtdbd^{2}q,
\]
where
\[
r_{\varepsilon}=\sqrt{b^{2}+q^{2}+\varepsilon^{2}},\quad\quad q^{2}= q_1^2+q_2^2.
\]
Adding and subtracting in the numerator the term $q^{2}a^{i}/2$, after some rearrangements this integral can be split into
\begin{equation}
{\cal F}^{i0}_\ve
(\varphi)=\int\!(G_{\varepsilon}^{i}+H_{\varepsilon}^{i})\,
\varphi(t+r_{\varepsilon},b\vec{v}+q_{a}\vec{N}_{a}+\vec{y})\,dtdbd^{2}q,\label{EBProof3}
\end{equation}
where
\begin{alignat}{1}
G_{\varepsilon}^{i}&=-\frac{e}{4\pi}\frac{\varepsilon^{2}(r_{\varepsilon}+b)a^i}
{\left(q^{2}+\varepsilon^{2}\right)^{2}},\label{gei}\\
H_{\varepsilon}^{i}&=\frac{e}{4\pi}\left(
\frac{\Pi_{ab}N_{a}^{j}N_{b}^{i}(r_{\varepsilon}
+b)^{2}a^{j}}{r_{\varepsilon}(q^{2}+\varepsilon^{2})^{2}}
-\frac{q^{2}a^i}{2r_{\varepsilon}(q^{2}
+\varepsilon^{2})}-\frac{\big(\vec{a}\cdot q_{c}\vec{N}_{c}\big)(r_{\varepsilon}+b)v^{i}}{r_{\varepsilon}(q^{2}+
\varepsilon^{2})}\right)\!,\label{hei}\\
\label{proj}
\Pi_{ab}&=q_{a}q_{b}-\frac{q^{2}}{2}\,\delta_{ab}.
\end{alignat}
To compute the limit
\be\label{fi0}
{\cal F}^{i0}
(\varphi)=\lim_{\varepsilon\rightarrow0}{\cal F}^{i0}_\ve(\varphi)
\ee
we will always exchange the limit with the
integral sign, resorting to the dominated convergence theorem. We will
treat the two terms in $\eqref{EBProof3}$ separately.

We start with the contribution involving $G_{\varepsilon}^{i}$,  performing the rescaling  ${q}_a\ra\varepsilon{q}_a$,
\be\label{ge}
\int\!G_{\varepsilon}^{i}\,
\varphi(t+r_{\varepsilon},b\vec{v}+q_{a}\vec{N}_{a}+\vec{y})\,dtdbd^{2}q=
-\frac{e}{4\pi}\int\!\!\frac{R_{\varepsilon}+b}
{(q^{2}+1)^{2}}\,a^{i}\varphi(t+R_{\varepsilon},b\vec{v}+\varepsilon q_{a}\vec{N}_{a}+\vec{y})\,dtdbd^{2}q,
\ee
where $R_{\varepsilon}=\sqrt{b^{2}+\varepsilon^{2}(q^{2}+1)}$. Since we have
\[
\lim_{\varepsilon\rightarrow0}\,(R_{\varepsilon}+b)=2bH(b),
\]
with $H$ the Heaviside function, the integral over $q_a$ in \eref{ge} becomes elementary and we obtain, see \eref{pi0},
\be
\lim_{\ve\ra0}\int\!G_{\varepsilon}^{i}\,
\varphi(t+r_{\varepsilon},b\vec{v}+q_{a}\vec{N}_{a}+\vec{y})\,dtdbd^{2}q=
-\frac{e}{2}\int \!bH(b)\,a^{i}\varphi(t+b,b\vec{v}+\vec{y})\,dtdb=
\label{pio}
\frac{1}{2}\,Q^{i0}(\varphi).
\ee
For what concerns, instead,  $H_{\varepsilon}^{i}$ one has the point-wise limit
\be\label{pw}
\lim_{V\ra 1} H_{\varepsilon}^{i}= H^i,
\ee
with $H^i$ given in \eref{hi}. It is straightforward to check that, as $\varepsilon\ra 0$, the second and
third terms of \eref{hei} lead in \eref{EBProof3} to convergent $q_a$-integrals. For what concerns the first term, adopting two-dimensional polar
coordinates $\left(q,\vartheta\right)$ instead of $\left(q_{1},q_{2}\right)$,
as $\varepsilon\ra 0$ the integral is asymptotic to $d^2q/q^2$ for $q\rightarrow0$ and seems, therefore, to produce a logarithmic divergence. However, if -- before taking the limit $\ve\ra 0$ -- we integrate {\it first} over $\vartheta$ and {\it then} over $q$, the traceless factor $\Pi_{ab}$ makes the integral convergent as $\ve\ra0$: for this reason it is said to be \textit{conditionally convergent}, that is,
convergent for a specific order of integrations. In conclusion the limit \eref{fi0} amounts to the expression $\eqref{EBProof5}$.

\subsection*{Appendix IV}

This appendix is devoted to the computation of the limit $\eqref{limK}$, whose result is \eref{KB1}, with $K^\m$ given in \eref{K}. We will suppose that the world-line $\vec y(t)$ is either bounded, or unbounded but fulfills the asymptotic conditions \eref{asym}, \eref{limUBv}. In both cases we can resort to the dominated convergence theorem to swap the limit $V\ra 1$ with the integral sign: in the first case this can be done thanks to the uniform bound $|\vec y(t)|<M$, and in the second thanks to the bound on the acceleration in \eref{limUBv} \[|\vec a(t)|=|\ddot{\vec \Delta}(t)|< \frac {c}{|t|^3}.\]

Applying to \eref{K} the same operations that led from $\eqref{CV}$ to $\eqref{VRegProof3}$, see in particular the change of variables \eref{bqVar}, we get
\[
{K}^{\mu}(\varphi)=-\frac{e(1-V^2)V^2}{2\pi}\int\!\frac{
\vec{a}\cdot q_{a}\vec{N}_{a}}{r(r-Vb)^{3}}\,\big(r,b\vec{v}+q_{d}\vec{N}_{d}\big)\,
\varphi\big(t+r,b\vec{v}+q_{c}\vec{N}_{c}+\vec{y}\big)\,dtdbd^{2}q,
\]
where the kinematical quantities $\vec y$, $\vec v$ and $\vec a$ are evaluated at time $Vt$ and $r=|\vec x|$. Performing the rescaling $q_{a}\rightarrow\sqrt{1-V^{2}}q_{a}$ we come to
\begin{align}
{K}^{\mu}(\varphi)=-\frac{eV^2}{2\pi\sqrt{1-V^{2}}}
\int&\frac{\vec{a}\cdot q_{a}\vec{N}_{a}(r+Vb)^{3}}{r(q^{2}+b^{2})^{3}}
\,\big(r,b\vec{v}+\sqrt{1-V^{2}}q_{d}\vec{N}_{d}\big)\cdot\nn\\
&\varphi\big({t}+r,b\vec{v}+\sqrt{1-V^{2}}
q_{c}\vec{N}_{c}+\vec{y}\big)\,dtdbd^{2}q,\label{KProof1}
\end{align}
where from now on $r=\sqrt{b^2+(1-V^2)q^2}$.
Setting
\[
I\equiv-\frac{eV^2}{2\pi\sqrt{1-V^{2}}}
\frac{\vec{a}\cdot q_{a}\vec{N}_{a}(r+Vb)^{3}}{r(q^{2}+b^{2})^{3}},
\]
and adding and subtracting from $\eqref{KProof1}$ the same term, we can write
\begin{alignat}{1}
{K}^{\mu}(\varphi) & =\int \! I\cdot\big(r,b\vec{v}+\sqrt{1-V^{2}}q_{a}\vec{N}_{a}\big)\cdot\nonumber \\
 & \quad\quad\quad\left(\varphi\big({t}+r,b\vec{v}+\sqrt{1-V^{2}}q_{c}\vec{N}_{c}+\vec{y}\big)
-\varphi\big({t}+r,b\vec{v}+\vec{y}\big)\!\right)\!dtdbd^{2}q\label{KProof2}\\
 & +\int \! I\cdot\big(r,b\vec{v}+\sqrt{1-V^{2}}q_{a}\vec{N}_{a}\big)\,\varphi\big({t}+r,b\vec{v}+\vec{y}\big)
dtdbd^{2}q.\label{KProof3}
\end{alignat}

We evaluate first the limit for $V\ra 1$ of the term $\eqref{KProof3}$. The contribution proportional
to the four-vector $\left(r,b\vec{v}\right)$ drops out, since
by symmetric integration over $d^{2}q$ we have
\[
\int\frac{\vec{a}\cdot q_{a}\vec{N}_{a}(r+Vb)^3}{r(q^{2}+b^{2})^{3}}\,
\big(r,b\vec{v}\big)\varphi\big({t}+r,b\vec{v}+\vec{y}\big)\,dtdbd^{2}q=0.
\]
We are therefore left with the limit of the contribution proportional to $\big(0,\sqrt{1-V^{2}}q_{a}\vec{N}_{a}\big)$, that we call
\[
H^{\mu}\equiv\lim_{V\rightarrow1}\int \! I\cdot\big(0,\sqrt{1-V^{2}}q_{a}\vec{N}_{a}\big)\varphi\big({t}+r,b\vec{v}+\vec{y}\big)
\,dtdbd^{2}q.
\]
Swapping the limit $V\ra 1$ with the integral sign we get (using that $\lim_{V\ra 1}r=|b|$)
\begin{alignat*}{1}
H^{i} & =-\frac{e}{2\pi}\lim_{V\rightarrow1}\int\frac{a^{j}q_{a}N_{a}^{j}\,(r+Vb)^{3}}
{r(q^{2}+b^{2})^{3}}\,
q_{c}N_{c}^{i}\,\varphi\big({t}+r,b\vec{v}+\vec{y}\big)\,dtdbd^{2}q\\
 &=-\frac{e}{2\pi}\int_{0}^{\infty}\!db\int\frac{a^{j}q_{a}N_{a}^{j}\left(2b\right)^{3}}
{b(q^{2}+b^{2})^{3}}\,q_{c}N_{c}^{i}\,\varphi\big(t+b,b\vec{v}+\vec{y}\big)\,dtd^{2}q\\
 & =-\frac{4e}{\pi}\int_{0}^{\infty}\!db\int\!\frac{b^2q_{a}q_{c}}{(q^{2}+b^2)^{3}}\,
a^{j}N_{a}^{j}N_{c}^{i}
\,\varphi\big(t+b,b\vec{v}+\vec{y}\big)\,dtd^{2}q.
\end{alignat*}
Integrating over $d^{2}q$, and using
\begin{equation}
\int\frac{q_{a}q_{c}\,d^{2}q}{(q^{2}+b^2)^{3}}=\frac{\pi}{4b^2}\,\delta_{ac},\quad \quad N_{a}^{i}N_{a}^{j}=\delta^{ij}-v^{i}v^{j},\quad\quad\vec{a}\cdot\vec{v}=0,\label{qId}
\end{equation}
through the shift $t\rightarrow t-b$ we get
\be\label{hii}
H^{i} =-e\int_{0}^{\infty}\!db\int\! a^{i}(t-b)\,\varphi\big(t,b\vec{v}(t-b)+\vec{y}(t-b)\big)\,dt.
\ee
Applying the limit $V\ra 1$ to \eref{KProof2}, \eref{KProof3} we get then, swapping in \eref{KProof2} again the limit with the integral sign
\begin{alignat}{1}
\mathcal{K}^{\mu}(\varphi)=\lim_{V\ra1}K^\m(\vp)&=  H^{\mu}-\frac{4e}{\pi}\!\int_{0}^{\infty}\!db\!\int\frac{a^{j}q_{a}N_{a}^{j}\,b^{3}}
{(q^{2}+b^{2})^{3}}\,
(1,\vec{v})\,q_{c}N_{c}^{i}\,\partial_{i}\varphi\big(t+b,b\vec{v}+\vec{y}\big)
\,dtd^{2}q\nonumber \\
&=H^{\mu}-e\!\int_{0}^{\infty}\!db\!\int\!(1,\vec{v})\,ba^{i}\partial_{i}\varphi\big(t+b,b\vec{v}
+\vec{y}\big)\,dt\nonumber \\
 & =H^{\mu}-e\!\int_{0}^{\infty}\!db\!\int\!(1,\vec{v}(t-b))\,ba^{i}(t-b)\,\partial_{i}
\varphi\big(t,b\vec{v}(t-b)\!+\!\vec{y}(t-b)\big)dt.\nn
\end{alignat}
We insert now \eref{hii} and notice that -- omitting the arguments of the test function -- we have
\be\label{KB0}
-ba^{i}(t-b)\,\partial_{i}\varphi=\frac{\pa\vp}{\pa b}.
\ee
In this way we obtain eventually
\begin{align}
\mathcal{K}^{\mu}(\varphi)&=e\int_{0}^{\infty}\!db\int\!\left(\frac{\pa}{\pa b},-\vec a(t-b)+\vec{v}(t-b)\frac{\pa}{\pa b}\right)\!\varphi\big(t,b\vec{v}(t-b)+\vec{y}(t-b)\big)dt\nn\\
&=e\!\int \! v^{\mu}(t-b)\,\varphi\big(t,b\vec{v}(t-b)+\vec{y}(t-b)\big)\,dt
\Big|_{b=0}^{b=\infty}.\label{KB2}
\end{align}

\subsection*{Appendix V}

In this appendix we  prove that in the integral \eref{A1}, that we write as
\[
A^\m_1(\vp)=\frac{e}{4\pi V}\!\int\!\big(f^\m_V(x)+g^\m_V(x)\big)\,d^4x,
\]
with
\begin{align}
f^\m_V(x)&=\frac{W^{\mu}(t)-\mathcal{V}_{\infty}^{\mu}}{r}\,
\varphi\big(r+ t/V,\vec{x}+\vec{y}(t)\big),\nn\\
g^\m_V(x)&= \frac{\mathcal{V}_{\infty}^{\mu}}{r}\,\big(
\varphi\big(r+ t/V,\vec{x}+\vec{y}(t)\big)-
\varphi\big(r+t/V,\vec{x}+\vec{v}_{\infty}t\big)\big),\label{gmu}\\
W^\m(t)&=(1,V\vec v(t)),\label{wm}
 \end{align}
we can swap the limit $V\ra 1$ with the integral sign, if $\vec y(t)$ satisfies the asymptotic conditions \eref{asym}, \eref{limUBv}. To do this we resort to the dominated convergence theorem, that requires to find {\it integrable} functions that dominate $f^\m_V$ and $g^\m_V$ {\it uniformly} in $V$. Since it is straightforward to find such functions in the integration region $r=|\vec x|<1$, we consider only the region $r>1$ and have thus $1/r<1$.

We consider first $f^\m_V$. Recalling \eref{asym}, \eref{aasym} and \eref{wm}, we perform in the corresponding integral the shift $\vec x\ra \vec x-\vec y(t)$ and, calling the new variable again $\vec x$, we have for each $\m$\footnote{Actually $f^0_V(x)$ is identically zero.}
\be\label{fmm}
\big|f^\m_V(x)\big|\le\big|\dot{\vec \Delta}(t)\big|
\left|\varphi\!\left(\sqrt{|\vec x-\vec y(t)|^2}+ t/V,\vec{x}\right)\right|\le
\frac{\big|\dot{\vec \Delta}(t)\big|}{\big(1+|\vec x|^2\big)^2}\,\Vert\vp\Vert,
\ee
where we multiplied the numerator as well as the denominator with ${\big(1+|\vec x|^2\big)^2}$, and $\Vert\vp\Vert$ is a combination of semi-norms. Thanks to the second bound in \eref{limUBv} the last function in \eref{fmm} is integrable (as well as $V$-independent).

For what concerns $g^\m_V$ we distinguish the regions $t>0$ and $t<0$ and use that for each $\m$ we have  $|{\cal V}^\m_\infty|\le1$. For $t>0$ we multiply the numerator as well as the denominator of both terms of $g_V^\m$ with $(r+t/V)^6$, obtaining
\[
\big|g^\m_V(x)\big|_{t>0}\le \frac{\Vert\vp\Vert}{(r+t/V)^6}\le \frac{\Vert\vp\Vert}{(r^2+t^2)^3},
\]
where $\Vert\vp\Vert$ indicates again a combination of semi-norms. The last function is, once more, integrable and $V$-independent.

In the region $t<0$ we perform in both terms of \eref{gmu} the shift $\vec x \ra \vec x-\vec v_\infty t$ and resort then to the {\it mean value theorem} to write ($R\equiv\sqrt{|\vec x-\vec v_\infty t|^2}$)
\[
g^\m_V(x)=\frac{{\cal V}^\m_\infty}{R} \, \vec{\Delta}(t)\cdot\vec{\nabla}\varphi\big(R+t/V,\vec{x}+\lambda\vec{\Delta}(t)\big),
\]
where $0\le\la\le1$. Multiplying and dividing this expression by
$\big(1+|\vec x+\la \vec\Delta(t)|^2\big)^2$ and remembering that $r=R>1$, we obtain the estimate
\[
\big|g^\m_V(x)\big|_{t<0}\le  \frac{|\vec\Delta(t)|\cdot\!{\Vert\vp\Vert}}{\big(1+|\vec x+\la \vec\Delta(t)|^2\big)^2},
\]
where $\Vert\vp\Vert$ is now a combination of semi-norms involving also derivatives of $\vp$. The first bound in \eref{limUBv} ensures that for $t<0$ the ``trajectory'' $\vec \Delta(t)$ is bounded and we have thus $|\lambda\vec \Delta(t)|<M$, for some constant $M$.
Proceeding as in \eref{ABDistProof1} and \eref{aint} we get therefore the uniform estimate
\[
\big|g^\m_V(x)\big|_{t<0}\le  \frac{|\vec\Delta(t)|\cdot\!{\Vert\vp\Vert}}{\big(1+(|\vec x|-M)^2H(|\vec x|-M)\big)^2}.
\]
Thanks to the first bound in \eref{limUBv} the r.h.s of this estimate is integrable.

\subsection*{Appendix VI}

In this appendix we rederive the field \eref{FB} -- produced by a bounded motion -- performing the distributional limit for $V\ra 1$ of the radiation  field  \eref{RV}, relying on the Li\'enard-Wiechert regularization \eref{YVB}.

Using \eref{fxe} and proceeding as in \eref{feps} we get
\be\nn
R^{\mu\nu}(\varphi)=\frac{e}{4\pi} \!\int \frac{X^{\mu}((UX)W^{\nu}-(WX)U^{\nu})}{r(UX)^{2}}\,\vp(X+Y) \,d^3x\,d\la -(\mu\leftrightarrow\nu),
\ee
where $X^\m=(r,\vec x)$ and $r=|\vec x|$ and the world-line $Y^\m(\la)$ is given by \eref{YVUR} or, equivalently, by \eref{YVB}. Considering again the electric field we obtain
\begin{equation}
R^{i0}(\varphi)=\frac{eV^2}{4\pi}\!\int\frac{(\vec{a}\cdot\vec{x})
(x^i-rV v^{i})-r(r-V\vec{v}\cdot\vec{x})\,a^{i}}
{r(r-V\vec{x}\cdot\vec{v})^{2}}\,
\varphi(t+r,\vec{x}+\vec{y})\,d^3x\,dt,
\end{equation}
where the kinematical variables $\vec y$, $\vec v$ and $\vec a$ are evaluated at time $Vt$. With  the change of variables \eref{bqVar} this expression can be brought to the form
\begin{equation}
R^{i0}
(\varphi)=\int\!(G_V^{i}+H_V^{i})\,
\varphi(t+r,b\vec{v}+q_{a}\vec{N}_{a}+\vec{y})\,dtdbd^{2}q,\label{ri0}
\end{equation}
where
\begin{alignat}{1}
G_V^{i}&=-\frac{eV^2}{4\pi}\,\frac{(1-V^2)(r+Vb)b^2a^i}
{\left(q^{2}+(1-V^2)b^2\right)^{2}},
\nn\\
H_V^{i}&=\frac{eV^2}{4\pi}\left(
\frac{\Pi_{ab}N_{a}^{j}N_{b}^{i}a^{j}}{r(r-Vb)^{2}}
-\frac{q^{2}a^i}{2r(q^{2}
+(1-V^2)b^2)}-\frac{\big(\vec{a}\cdot q_{c}\vec{N}_{c}\big)(Vr-b)v^{i}}{r(r-Vb)^2}\right)\!,\nn
\end{alignat}
and $\Pi_{ab}$ is given in \eref{proj}. Taking advantage of the similarities between these formulae and \eref{EBProof3}-\eref{hei}, we can proceed in the same manner as after \eref{ge}. In the integral containing $H_V^i$ in \eref{ri0} we can take the limit $V\ra 1$ under the integral sign, the resulting (conditionally convergent) integral giving rise to the second term in \eref{EBProof5}. In the integral involving $G_V^i$ we perform the rescaling $q_a\ra \sqrt{1-V^2}q_a$ -- instead of $q_a\ra \ve q_a$ -- and send then $V\ra 1$, the result being again \eref{pio}. From \eref{ri0} we obtain hence
\[
\lim_{V\ra 1}R^{i0}(\varphi)=\frac{1}{2}\,Q^{i0}(\vp)+\mathcal{P}(\mathcal{F}^{i0}_{reg})(\vp).
\]
The same calculation for ${ R}^{ij}$ leads to an analogous result and we retrieve thus \eref{rmn}.

\subsection*{Appendix VII}

We prove here that for a particle moving on a trajectory $\vec y(t)$ that in the infinite past becomes linear (see \eref{asym} and \eref{limUBv}) at each time $t$ the regular contribution of the total field \eref{Fug}, {\it i.e.} the field  $\mathcal{F}^{\mu\nu}_{reg}$ in \eref{RBpointWise}, vanishes ``in front of'' the shock-wave \eref{csw}, that is it vanishes for any $\vec x$ such that $\vec x\cdot\vec v_\infty>t$. To this order we must show that the retarded-time equation
\eref{llcaus} for these $\vec x$ does not admit a solution for $\la$.

We begin writing the trajectory \eref{asym}, \eref{limUBv}  in the form
\begin{equation}
\vec{y}(t)=(t-\varepsilon(t))\,\vec{v}_{\infty}+\vec{T}(t),\label{NoSolProof3}
\end{equation}
where $\vec{T}(t)\perp\vec{v}_{\infty}$ and
\be\label{limte}
\lim_{t\rightarrow-\infty}\vec{T}(t)=0,\quad\quad\lim_{t\rightarrow-\infty}\varepsilon(t)=0.
\ee
Since $|\vec v(t)|=|d\vec y(t)/dt|=1$ we have, moreover, $\varepsilon(t)\geq0$. This means in particular that the particle at each time $t$ stays always behind the shock-wave, located  on the plane  $\vec x\cdot\vec v_\infty=t$.

Parameterizing the trajectory with time, $y^0(\la)=\la$, we write the conditions \eref{llcaus} in the form
 \begin{equation}
t-\la=|\vec x-\vec{y}(\la)|\label{NoSolProof2}
\end{equation}
and choose a generic point $\vec x$  {\it in front of} the shock-wave:
\begin{equation}
\vec{x}=(t+l)\vec{v}_{\infty}+\vec{N},\quad \quad l>0,\quad\quad\vec{N}\cdot\vec{v}_{\infty}=0.\label{NoSolProof1}
\end{equation}
Equation \eref{NoSolProof2} becomes then
\be
t-\la =\sqrt{\big(t-\la+l+\varepsilon(\la)\big)^{2}+\big(\vec{N}+\vec{T}(\la)\big)^{2}}
\label{NoSolProof4}
\, \,>|t-\la|,
\ee
and has thus no solution.

Considering, on the other hand, a point $\vec x$ {\it behind} the shock-wave -- given by \eref{NoSolProof1} with $l<0$ -- equation \eref{NoSolProof2} takes again the form \eref{NoSolProof4}, that can be written also as
\[
f(\la)\equiv  2(t-\la)L-L^{2}-\big(\vec{N}+\vec{T}(\la)\big)^2=0,
\]
where we have set $L\equiv -l-\varepsilon(\la)$. Since we have $f(t)<0$ and (thanks to $\lim_{\la\ra-\infty}L=-l>0$ and to \eref{limte}) $\lim_{\la\ra-\infty}f(\la)=+\infty$, due to continuity there exists a $\la_0<t$ such that $f(\la_0)=0$. This means that {\it behind} the shock-wave the field $\mathcal{F}^{\mu\nu}_{reg}(t,\vec x)$ is generically  non-vanishing.

\subsection*{Appendix VIII}

In this appendix we furnish an independent proof that the regular field  $\mathcal{P}(\mathcal{F}^{\mu\nu}_{reg})$ satisfies for bounded trajectories the equation \eref{demo2}, and for unbounded ones the equation \eref{curd2}. Thanks to manifest Lorentz-invariance we may restrict ourselves to the time-component of these equations --
respectively $ \vec\nabla\!\cdot\!\vec { E}_{reg}=\mathcal{J}^0/2$ and  $\vec\nabla\!\cdot\!\vec { E}_{reg}=(\mathcal{J}^0-\mathcal{J}^0_{\mathcal L})/2$ --
involving only the electric field $E_{reg}^i= \mathcal{P}(\mathcal{F}^{i0}_{reg})$ given by the second term in \eref{EBProof5}. We write this field as
\[
 {E}_{reg}^i(\vp)=\int \!H^i(b,q,t) \,\vp(t+r,\vec x+\vec y(t))\, dbd^2qdt,
\]
where we may look at the function $H^i$ \eref{hi} also as a function of $\vec x= b\vec v + q_c \vec N_c$ (see \eref{bqVar}) and $t$; in this case we have $dbd^2q=d^3x$. Away from the singularity-string (and hence also away from the position of the particle) {\it i.e.} at $q^a\neq 0$, this function obeys the identities
\be
\label{trans}
x^i H^i=0,\quad\quad \pa_iH^i=0.
\ee
The first identity follows by inspection from \eref{hi} and represents the standard result that the radiation field -- of a massive as well as of massless particle -- is always orthogonal to the ``retarded'' radial direction. The second identity follows from the fact that away from the singularity-string we have $\vec\nabla\!\cdot\!\vec{ E}_{reg}=0$; the dependence of $\vp$ on $\vec x$ through the time variable $t+r$ has no effect with this respect, since $\pa_i (t+r) =x^i/r$ and $x^iH^i=0$.

To evaluate $\vec\nabla\!\cdot\!\vec{ E}_{reg}$ we must apply it to a test function
\[
\big(\vec\nabla\!\cdot\!\vec{ E}_{reg}\big)(\vp)=-{E}_{reg}^i(\pa_i\vp)=
-\int \!H^i(\vec x,t) \,\pa_i \vp(t+r,\vec x+\vec y(t))\,d^3xdt.
\]
Due to the singularities of $H^i$ at $q^2= r^2-(\vec v\cdot\vec x)^2=0$ we are not allowed to integrate the derivatives $\pa_i$ by parts. We may do so, however, if we insert the step function  ($H$ denotes the Heaviside function)
\[\Theta_\ve(\vec x)\equiv  H\big( r^2-(\vec v\cdot\vec x)^2-\ve ^2\big),\]
excluding a tubular neighborhood around the singularity-string, and take eventually the limit $\ve\ra 0$
\be\label{dee}
\big(\vec\nabla\!\cdot\!\vec{ E}_{reg}\big)(\vp)=-\lim_{\ve\ra 0}
\int \!\Theta_\ve(\vec x)\, H^i(\vec x,t) \,\pa_i \vp(t+r,\vec x+\vec y(t))\,d^3xdt.
\ee
In integrating by parts, thanks to the identities \eref{trans}, only the step function contributes,
\[
\pa_i \Theta_\ve(\vec x)= 2\big(x^i- (\vec v\cdot\vec x)v^i \big)  \delta\big(r^2-(\vec v\cdot\vec x)^2-\ve ^2\big),
\]
where the term proportional to $x^i$ drops out from \eref{dee}.
Switching back to the coordinates $(b,q_a)$ we get therefore
\[
\big(\vec\nabla\!\cdot\!\vec{ E}_{reg}\big)(\vp)=
-2\lim_{\ve\ra 0}\int \!b v^i \delta(q^2-\ve^2)\,H^i(b,q,t)\,\vp(t+r,\vec y+b\vec v+q_c\vec N_c)\,dbd^2qdt,
\]
where it is understood that all kinematical variables are evaluated at time $t$. Looking at \eref{hi} one sees that only the third term of $H^i$ has a non-vanishing projection along $v^i$ and, thanks to $\delta(q^2-\ve^2)=\delta(q-\ve)/2\ve$,
one arrives at
\be\label{dive1}
\big(\vec\nabla\!\cdot\!\vec{ E}_{reg}\big)(\vp)=
\frac{e}{4\pi}
\lim_{\ve\ra 0}\int\! \delta(q-\ve)\,\frac{b\,(r+b)\big(q_d\vec N_d\cdot \vec a\big)}{r\ve^3} \, \vp(t+r,\vec y+b\vec v+q_c\vec N_c)\,dbd^2qdt,
\ee
where now $r=\sqrt{b^2+\ve^2}$.
Being conditionally convergent, see Section \ref{lof}, the integral must be evaluated integrating first over the polar angle of $q_a$. This implies that, as $\ve\ra 0$, in expanding $\vp$ in powers of $ q\equiv\ve $ it is only the first-order term that contributes,
\[
\vp(t+r,\vec y+b\vec v+q_c\vec N_c)= \vp(t+|b|,\vec y+b\vec v) + q_c N_c^j  \pa_j  \vp(t+|b|,\vec y+b\vec v)+o(q^2).
\]
Taking into account that $\lim_{\ve\ra 0}(r+b)=2bH(b)$ and using \eref{bq}, through symmetric integration over $q_a$ from \eref{dive1} one obtains finally (performing the shift $t\ra t-b$  and applying the identity \eref{KB0})
\begin{align}
\nn
\big(\vec\nabla\!\cdot\!\vec{ E}_{reg}\big)(\vp)&=\frac {e}{2}\int\!\left( \int_0^\infty \!ba^i(t-b)\pa_i \vp(t,\vec y(t-b)+b\vec v(t-b))\,db\right)dt\\
&=\frac {e}{2}\int\!\vp(t,\vec y(t))\,dt-\frac {e}{2}\int\!\left(\lim_{b\ra\infty}\vp(t,\vec y(t-b)+b\vec v(t-b))\!\right)dt.\label{check}
\end{align}
The first term equals $\mathcal{J}^0(\vp)/2$. The second term is zero for bounded trajectories, while it equals $-\mathcal{J}^0_{\cal L}(\vp)/2$ for unbounded ones,
see \eref{limgammaUB}.  Equation \eref{check} reduces thus just to the $\n=0$ components of \eref{demo2} and \eref{curd2}.


\begin{thebibliography}{1}

\bibitem{K} T. Kinoshita, \textit{Mass singularities of Feynman amplitudes},
J. Math. Phys. \textbf{3} (1962) 650.

\bibitem{LN} T.D. Lee and M. Nauenberg, \textit{Degenerate systems
and mass singularities}, Phys. Rev. \textbf{133} (1964) B1549.

\bibitem{LM} M. Lavelle and D. McMullan, \textit{Collinearity, convergence
and cancelling infrared divergences}, JHEP \textbf{03} (2006) 026,
[arXiv:hep-ph/0511314].

\bibitem{W} S. Weinberg, \textit{The Quantum Theory of Fields I},
Cambridge University Press, Cambridge, (1995), pg. 552.

\bibitem{RR} I. Robinson and K. Rozga, \textit{Lightlike contractions
on Minkowski space-time}, J. Math. Phys. \textbf{25} (1984) 499.

\bibitem{AE} P. Aichelburg and F. Embacher, \textit{Lightlike contractions
in curved space-time}, in ``Gravitation and Geometry'', eds. W.
Rindler and A. Trautman, Bibliopolis, edizioni di filosofia e scienze,
Naples (1987).

\bibitem{JKO} R. Jackiw, D. Kabat and M. Ortiz, \textit{Electromagnetic
fields of a massless particle and the eikonal}, Phys. Lett. \textbf{B277}
(1992) 148, [arXiv:hep-th/9112020].

\bibitem{AL} F. Azzurli and K. Lechner, \textit{The Li\'enard-Wiechert
field of accelerated massless charges}, Phys. Lett. \textbf{A377}
(2013) 1025, [arXiv:1212.3532[hep-th]].

\bibitem{B} A.J. Baltz, \textit{Coulomb potential from a particle in uniform ultrarelativistic motion}, Phys. Rev. \textbf{A52} (1995) 4970.

\bibitem{Y} Y. Yaremko, \textit{Radiation reaction and renormalization for a photon-like charged particle}, Electron. J. Theor. Phys. \textbf{9} (2012) 153, [arXiv:0907.3694[math-ph]].

\bibitem{LPAM} K. Lechner and P.A. Marchetti, \textit{Variational
principle and energy-momentum tensor for relativistic Electrodynamics
of point charges}, Ann. Phys. \textbf{322} (2007) 1162, [arXiv:hep-th/0602224].


\end{thebibliography}
\end{document}